\documentstyle[preprint,eqsecnum,aps]{revtex}

\begin{document}
\draft
\title{{\Large Aharonov-Bohm Effect in Cyclotron and Synchrotron Radiations}}
\author{V.G. Bagrov\thanks{%
On leave from Tomsk State University and Tomsk Institute of High Current
Electronics, Russia}, D.M. Gitman\thanks{%
e-mail: gitman@fma.if.usp.br}, A. Levin\thanks{%
School of Physics and Astronomy, University of Nottingham, UK}, and V.B.
Tlyachev\thanks{%
Tomsk Institute of High Current Electronics, Russia}}
\address{Instituto de F\'{\i}sica, Universidade de S\~ao Paulo,\\
C.P. 66318, 05315-970 S\~ao Paulo, SP, Brasil}
\date{\today}
\maketitle

\begin{abstract}
We study the impact of Aharonov-Bohm solenoid on the radiation of a charged
particle moving in a constant uniform magnetic field. With this aim in view,
exact solutions of Klein-Gordon and Dirac equations are found in the
magnetic-solenoid field. Using such solutions, we calculate exactly all the
characteristics of one-photon spontaneous radiation both for spinless and
spinning particle. Considering non-relativistic and relativistic
approximations, we analyze cyclotron and synchrotron radiations in detail.
Radiation peculiarities caused by the presence of the solenoid may be
considered as a manifestation of Aharonov-Bohm effect in the radiation. In
particular, it is shown that new spectral lines appear in the radiation
spectrum. Due to angular distribution peculiarities of the radiation
intensity, these lines can in principle be isolated from basic cyclotron and
synchrotron radiation spectra.
\end{abstract}

\section{Introduction}

Aharonov-Bohm (AB) effect \cite{AhaBo59} plays an important role in quantum
theory refining the status of electromagnetic potentials in this theory.
First this effect was discussed in relation to a study of interaction
between a non-relativistic charged particle and an infinitely long and
infinitesimally thin magnetic solenoid field\footnote{%
A similar effect was discussed earlier by Ehrenberg and Siday \cite{EhrSi49}}
(further AB field). It was discovered that particle wave functions vanish at
the solenoid line. In spite of the fact that the magnetic field vanishes out
of the solenoid, the phase shift in the wave functions is proportional to
the corresponding magnetic flux \cite{WuYa75}. A non-trivial particle
scattering by the solenoid is interpreted as a possibility for quantum
particles to ''feel'' potentials of the corresponding electromagnetic field.
Indeed, potentials of AB field do not vanish out of the solenoid. A number
of theoretical works and convinced experiments was done to clarify and prove
the existence of AB effect. The detailed exposition of this activity can be
encountered, for example, in \cite{OlaPo85,Skarz86,PesTo89,Nambu98}. In
particular, it was shown \cite{SerSk88,GalVo90,SkaAuJ96} that AB scattering
is accompanied by an electromagnetic radiation. Pair creation by a photon in
the presence of AB field was calculated in \cite{SkaAuJ96}. The interaction
between electron spin and AB field leads to Dirac wave functions that do not
vanish at the solenoid line. Thus, the issue of spin changes slightly the
interpretation of AB effect. Theoretical study of AB scattering for spinning
particles was presented in many papers, see for example \cite{Hagen91} and 
\cite{CouPe93}. AB effect was also discussed in connection with fractional
spin and statistics \cite{Wilcz82} and with cosmic strings \cite
{Gerbe89,AlfWi89}. AB effect in anyon scattering was considered in \cite
{GomSi98}; radiative corrections to the effect were calculated in \cite
{AlbGoS00}. AB scattering within the Chern-Simons theory of scalar particles
was studied in \cite{BozFaP95}. There exist impressive applications of AB
effect in solid state physics \cite{Bergm84,LeeRa85}.

A splitting of Landau levels in a superposition of parallel uniform magnetic
field and AB field (further magnetic-solenoid field) gives an example of AB
effect for bound states. First, exact solutions of Schr\"{o}dinger equation
in the magnetic-solenoid field (non-relativistic case) were studied in \cite
{Lewis83}. Then these solutions were used in \cite{Sereb85,BagGiS86,SerSk89}
to discuss AB effect.

It is well-known that a charged particle irradiates moving in a uniform
magnetic field. The corresponding radiation is called cyclotron one (CR) in
the non-relativistic case; it is called synchrotron radiation (SR) in the
relativistic case. In the present article we study how the presence of AB
field affects CR and SR. It is clear that classical trajectories do not feel
the presence of AB field whenever they do not intersect the solenoid. Thus,
from the classical point of view, CR and SR are not affected by the presence
of AB field. However, the latter field changes quantum trajectories, thus we
expect that in the framework of quantum theory characteristics of CR and SR
may be affected by such a field. We calculate spontaneous one-photon
radiation of a particle (both spinless and spinning) in the
magnetic-solenoid field in the framework of quantum theory. We consider from
the beginning quantum relativistic problem in order to analyze consistently
both relativistic (SR) and non-relativistic (CR) cases. One ought to mention
that conventional CR and SR were studied in detail in numerous works (see 
\cite{SokTe68} and Refs. there). The analysis of the radiation in the
magnetic-solenoid field is much more complicated and contains many new
aspects and technical details.

The article is organized in the following way: In Sect.II we present exact
solutions of Klein-Gordon and Dirac equations in the magnetic-solenoid field
and analyze the energy spectrum of particles in such a field. In Sect.III
matrix elements of transitions (both for spinless and spinning particles)
with one-photon radiation are calculated exactly. In Sect.IV we analyze
frequencies of the radiation. Spinless particle radiation is studied in
detail in Sect.V. Here an exact expression for the radiation\ intensity is
obtained. The non-relativistic approximation, semiclassical approximation,
and weak magnetic field limit are considered. Besides, we reveal some
important peculiarities of angular distribution of the radiation. Results
that were obtained for spinless particle radiation are generalized to
spinning particle case in Sect.VI. Particular emphasis is given to electron
transitions that cause low frequency (less that the basic synchrotron
frequency) radiation. In the end, we summarize results focusing our
attention on manifestations of AB effect in CR and SR.

\section{Relativistic particle in magnetic-solenoid field}

As was mentioned above, the magnetic-solenoid field is a superposition of a
constant uniform magnetic field of strength $H$ directed along the axis $z$
and a solenoid field (AB field). The latter field is created by an
infinitely long and infinitesimally thin solenoid situated along the same
axis $z.$ The solenoid creates a finite magnetic flux $\Phi $ along the axis 
$z$. The magnetic-solenoid field is given by electromagnetic potentials of
the form 
\begin{eqnarray}
A_{1} &=&x^{2}\left( \frac{\Phi }{2\pi r^{2}}+\frac{H}{2}\right)
,\;A_{2}=-x^{1}\left( \frac{\Phi }{2\pi r^{2}}+\frac{H}{2}\right)
,\;A_{0}=A_{3}=0,  \nonumber \\
\;\;r^{2} &=&(x^{1})^{2}+\left( x^{2}\right) ^{2},\;x^{0}=ct,\;{\bf x=(}%
x^{i}),\;x^{1}=x,\;x^{2}=y,\;x^{3}=z\,.  \label{1}
\end{eqnarray}
The potentials (\ref{1}) define the magnetic field ${\bf H\;}$of the form$%
{\bf \ }$ 
\begin{equation}
{\bf H}=(0,0,H_{z}),\;H_{z}=H+\Phi \delta (x^{1})\delta (x^{2})\;.  \label{3}
\end{equation}
It is convenient to present the magnetic flux $\Phi \;$as 
\begin{equation}
\Phi =(l_{0}+\mu )\Phi _{0},\quad \Phi _{0}=2\pi c\hbar /\left| e\right|
\;,\;0\leq \mu <1,\;l_{0}\in Z.  \label{2}
\end{equation}
The integer $l_{0}$ gives a number of quanta $\Phi _{0}$ in the total flux $%
\Phi $. The quantity $\mu $\ will be called the mantissa of the magnetic
flux $\Phi .\;$In the cylindrical coordinates $r,\varphi ,$%
\begin{equation}
x^{1}=r\cos \varphi ,\;x^{2}=r\sin \varphi ,\;\rho =\frac{\gamma r^{2}}{2}%
,\;\gamma ={\frac{|{eH|}}{{c\hbar }}}>0\,,  \label{4}
\end{equation}
the non zero potentials have the form 
\begin{equation}
\frac{|e|}{c\hbar }A_{1}=\frac{l_{0}+\mu +\rho }{r}\sin \varphi \,,\;\frac{%
|e|}{c\hbar }A_{2}=-\frac{l_{0}+\mu +\rho }{r}\cos \varphi \,\,.  \label{6}
\end{equation}
Doing a transformation of relativistic wave functions $\Psi
(x)=e^{-il_{0}\varphi }\tilde{\Psi}(x)\,,$ we can eliminate $l_{0}$
dependence from the potentials. Indeed, electromagnetic potentials enter in
relativistic wave equations via operators of momenta $\hat{P}_{\mu }=i\hbar
\partial _{\mu }-\frac{e}{c}A_{\mu }\,$. Thus, equations for $\tilde{\Psi}(x)
$ contain momentum operators of the form 
\begin{eqnarray}
&&e^{il_{0}\varphi }P_{\mu }e^{-il_{0}\varphi }=\hbar \left( i\partial _{\mu
}+\bar{A}_{\mu }\right) ,\;\bar{A}_{0}=\bar{A}_{3}=0\,,  \nonumber \\
&&\bar{A}_{1}=\frac{\mu +\rho }{r}\sin \varphi ,\;\bar{A}_{2}=-\frac{\mu
+\rho }{r}\cos \varphi \,.  \label{8}
\end{eqnarray}
Therefore, the functions $\tilde{\Psi}(x)$ depend on the mantissa of the
magnetic flux only.

Consider first solutions of Klein-Gordon equation 
\begin{equation}
\left( \hat{P}^{2}-m_{0}^{2}c^{2}\right) \Psi (x)=0\,,\;\hat{P}_{\mu
}=i\hbar \partial _{\mu }-\frac{e}{c}A_{\mu }\,  \label{9}
\end{equation}
in the solenoid-magnetic field. The operators $\hat{P}_{0},\hat{P}_{3},$ and 
$\hat{L}_{z}=x^{2}p_{1}-x^{1}p_{2}\,=-i\hbar \partial _{\varphi }\;$are
integrals of motion in the case under consideration ( $\hat{L}$\ is angular
momentum operator). We are looking for solutions of (\ref{9}) that are
eigenvectors for these operators, 
\begin{equation}
\hat{P}_{0}\Psi =\hbar k_{0}\Psi ,\;\hat{P}_{3}\Psi =\hbar k_{3}\Psi ,\;\hat{%
L}_{z}\Psi =\hbar \left( l-l_{0}\right) \Psi \,,\;l\in Z.  \label{10}
\end{equation}
The integer $l\;$is called the azimuthal quantum number. As a consequence of
(\ref{10}) and (\ref{9}), we have 
\begin{equation}
\hat{P}_{r}^{2}\Psi =\hbar ^{2}k^{2}\Psi ,\;\,\hat{P}_{r}^{2}=\hat{P}%
_{1}^{2}+\hat{P}_{2}^{2}\,,\;k_{0}^{2}=m^{2}+k_{3}^{2}+k^{2},\;m=\frac{m_{0}c%
}{\hbar }\,.  \label{14}
\end{equation}
Solutions of the equations (\ref{9}), (\ref{10}) can be written as 
\begin{equation}
\Psi \left( x\right) =e^{-i\Gamma }\psi \left( \rho \right) \,,\,\,\Gamma
=\,k_{0}x^{0}+k_{3}x^{3}+\,\left( l_{0}-l\right) \varphi \,,  \label{15}
\end{equation}
where the functions$\;\psi \left( \rho \right) \;$obey the equation 
\begin{equation}
\rho \psi ^{\prime \prime }+\psi ^{\prime }+\left[ \bar{n}+\frac{1}{2}-\frac{%
\left( \bar{l}+\rho \right) ^{2}}{4\rho }\right] \psi =0\,,\,k^{2}=2\gamma
\left( \bar{n}+\frac{1}{2}\right) ,\,\bar{l}=l+\mu \,.  \label{16}
\end{equation}
Solutions of this equation can be expressed via Laguerre functions$%
\;I_{n,m}(x).\;$The latter functions are defined (for any complex $%
n,m,x,\;n\neq -1,-2,-3,...)$ by the relation 
\begin{equation}
I_{n,m}(x)=\sqrt{\frac{{\Gamma (1+n)}}{{\Gamma (1+m)}}}\,{\frac{\exp (-x/2)}{%
{\Gamma (1+n-m)}}}x^{\frac{{n-m}}{2}}\Phi (-m,n-m+1;x)\,,  \label{17}
\end{equation}
where $\Phi \left( \alpha ,\beta ;x\right) \;$is the confluent
hypergeometric function (\cite{GraRy94}, 9.210).\ The Laguerre functions$%
\;I_{\alpha +m,m}(x)$ are quadratically integrable on the interval $x\geq 0$
whenever $m=0,1,2,...,$ and $%
\mathop{\rm Re}%
\alpha >-1.$ These functions form a complete and orthonormal set on the
interval $x\geq 0$ whenever $m=0,1,2,...,$ and $%
\mathop{\rm Im}%
\alpha =0,\;\alpha >-1.$ Namely, 
\begin{eqnarray}
\sum_{n=0}^{\infty }{I_{\alpha +n,n}(x)I_{\alpha +n,n}(y)} &=&{\delta }%
\left( x-y\right) ,\;x,y>0\,,  \label{18} \\
\int\limits_{0}^{\infty }{I_{\alpha +n,n}(x)I_{\alpha +m,m}(x)dx} &=&\delta
_{m,n},\;\alpha >-1,\;n,m=0,1,2,...\;.  \label{19}
\end{eqnarray}
It is a matter of direct verification (using an equation for the confluent
hypergeometric functions) to prove that a general solution of the
differential equation

\begin{equation}
4x^{2}I^{\prime \prime }+4xI^{\prime }-\left[ x^{2}-2x\left( 1+s+n\right)
+\left( s-n\right) ^{2}\right] I=0\,  \label{20}
\end{equation}
has the form $I=c_{1}I_{s,n}+c_{2}I_{n,s}\,.$\ The functions $I_{s,n},\;$and 
$I_{n,s}\,$are linearly independent for $s-n\notin Z.\;$Otherwise we have 
\begin{equation}
I_{n,s}=\left( -1\right) ^{n-s}I_{s,n}\,,\;s-n\in Z\,.  \label{22}
\end{equation}
Let $m$ be an integer and non-negative; then Laguerre functions are
connected to the Laguerre polynomials $L_{m}^{\alpha }(x)$ by the relation ( 
\cite{GraRy94}, 8.970) 
\begin{equation}
I_{\alpha +m,m}(x)=\sqrt{\frac{{\Gamma (1+m)}}{{\Gamma (1+\alpha +m)}}}\exp
(-x/2)x^{\frac{{\alpha }}{2}}L_{m}^{\alpha }(x)\;,\;\;\;m=0,1,2,...\;.
\label{23}
\end{equation}
Taking the above information into account, we can see that bounded and
quadratically integrable solutions of Eq.(\ref{16}) are divided in two
types, $\psi _{n,l}^{(j)}(r),\;j=1,2,\;$ 
\begin{eqnarray}
\psi _{n,l}^{(1)}(r) &=&I_{\bar{n},\bar{n}-\bar{l}}(\rho ),\;\bar{n}=n+\mu
,\;0\leq l\leq n,  \nonumber \\
\psi _{n,l}^{(2)}(r) &=&I_{\bar{n}-\bar{l},\bar{n}}(\rho ),\;\bar{n}%
=n,\;l<0,\;n\in Z\,.  \label{24}
\end{eqnarray}
The states of the first type ($j=1$) correspond to the energy spectrum of
the form 
\begin{equation}
k_{0}^{2}=m^{2}+k_{3}^{2}+2\gamma (n+\mu +{\frac{1}{2}}),\;0\leq l\leq n\,,
\label{25}
\end{equation}
and ones of the second type ($j=2$) correspond to the following spectrum 
\begin{equation}
k_{0}^{2}=m^{2}+k_{3}^{2}+2\gamma (n+{\frac{1}{2}}),\;l<0\,.  \label{26}
\end{equation}
The integer $n\geq 0$ is referred to as the principle quantum number. Note
that the spectrum (\ref{26}) of the second type states corresponds exactly
to the spectrum of spinless particles in a uniform magnetic filed. The
spectrum (\ref{25}) is deformed by the presence of the solenoind field
whenever $\mu \neq 0.$ Thus, the solenoind field partially lifts a
degeneracy of the magnetic field spectrum with respect to the quantum number 
$l$ whenever $\mu \neq 0.$ Namely, in the general case, the particle energy
spectrum in the magnetic-solenoid field depends on {\rm sign}$\,l\,.$

In accordance with Eq. (\ref{24}), it is convenient to define an effective
quantum number$\;\bar{n}$ by the relation 
\begin{equation}
\bar{n}=n+\mu \left( 2-j\right) =\left\{ 
\begin{array}{c}
n+\mu \,,\;j=1 \\ 
n\,,\;j=2,\;
\end{array}
\right. ,\;n=0,1,2,...\,\;.  \label{27}
\end{equation}
Then Eqs. (\ref{25}), (\ref{26}) can be integrated into a single formula 
\begin{equation}
k_{0}^{2}=m^{2}+k_{3}^{2}+2\gamma (\bar{n}+{\frac{1}{2}}),\;\bar{l}\leq \bar{%
n}.  \label{28}
\end{equation}

We stress that the solutions (\ref{24}) vanish at $r=0.$ That allows us to
speak about AB effect in the case under consideration whenever $\mu \neq 0.$

Similar to Klein-Gordon equation, the Dirac one (in the magnetic-solenoid
field) 
\begin{equation}
(\gamma ^{\mu }\hat{P}_{\mu }-m_{0}c)\Psi (x)=0  \label{29}
\end{equation}
admits $\hat{P}_{0},\,\hat{P}_{3}$ to be integrals of motion. Besides, $\hat{%
J}_{z}=\hat{L}_{z}+{\frac{\hbar }{2}}\Sigma _{3}\,\;(\hat{J}$ is the total
angular momentum operator and ${\bf \Sigma =}{\rm diag}\left( {\bf \sigma
,\sigma }\right) )$ is an integral of motion as well. Thus, we are looking
for solutions of (\ref{29}) that are eigenvectors for these integrals of
motion, 
\begin{equation}
\hat{P}_{0}\Psi =\hbar k_{0}\Psi ,\;\hat{P}_{3}\Psi =\hbar k_{3}\Psi ,\;\hat{%
J}_{z}\Psi =\hbar (l-l_{0}-{\frac{1}{2}})\Psi ,\;\;.  \label{31}
\end{equation}
Solutions of Eqs. (\ref{29}), (\ref{31}) can be written in the form 
\begin{equation}
\Psi (x)=N_{D}\exp (-i\Gamma )\left( 
\begin{array}{c}
e^{-i\varphi }c_{1}\psi _{n-1,l-1}^{(j)}\left( \rho \right) \\ 
ic_{2}\psi _{n,l}^{(j)}\left( \rho \right) \\ 
e^{-i\varphi }c_{3}\psi _{n-1,l-1}^{(j)}\left( \rho \right) \\ 
ic_{4}\psi _{n,l}^{(j)}\left( \rho \right)
\end{array}
\right) \,,  \label{32}
\end{equation}
where $N_{D}$ is a normalization factor and\ the constant bispinor $%
C=(c_{a},\,a=1,2,3,4)$ is subjected to the following algebraic system of
equations (we use the standard representation for $\gamma $-matrices) 
\begin{equation}
AC=0,\;A=\gamma ^{0}k_{0}+\gamma ^{3}k_{3}-\sqrt{2\gamma \bar{n}}\gamma
^{1}-m.  \label{33}
\end{equation}
The system (\ref{33}) has a nontrivial solution whenever 
\begin{equation}
\det A=\left( k_{0}^{2}-m^{2}-k_{3}^{2}-2\gamma \bar{n}\right) ^{2}=0\;.
\label{34}
\end{equation}
It follows from (\ref{34}) that the rank of the matrix $A$ equals $2$. Thus,
a nontrivial general solution of Eqs. (\ref{33}) contains two arbitrary
constants and can be written in the following block form via an arbitrary
spinor $\upsilon $, 
\begin{equation}
C=\left( 
\begin{array}{c}
\left( k_{0}+m\right) \upsilon \\ 
\left( \sqrt{2\gamma \bar{n}}\sigma _{1}-k_{3}\sigma _{3}\right) \upsilon
\end{array}
\right) ,\;C^{+}C=2k_{0}\left( k_{0}+m\right) \upsilon ^{+}\upsilon \,.
\label{35}
\end{equation}
As in the spinless particle case, we have here two types of states ($j=1,2$%
). The energy spectrum of spinning particles in the magnetic-solenoid field
follows from (\ref{34}), 
\begin{equation}
k_{0}^{2}=m^{2}+k_{3}^{2}+2\gamma \bar{n}\,.  \label{36}
\end{equation}

States of the second type (with $j=2)$ have one spin orientation only
whenever $n=0.$ Indeed, in such a case we must set $c_{1}=c_{3}=0.$ Thus, 
\begin{equation}
\upsilon =\left( 
\begin{array}{c}
0 \\ 
1
\end{array}
\right) ,\;n=0,\;j=2\,.  \label{37}
\end{equation}
In this case, the wave functions (\ref{32}) are eigenvectors for the
operator $\Sigma _{3}$ ($\Sigma _{3}\Psi =-\Psi $) with the eigenvalue $-1$
(the electron spin is always opposite to the magnetic field). That fact is
well-known \cite{SokTe68,BagGi90} in the absence of the solenoid field .

The states of the first type ($j=1$) vanish at $r=0$ whenever $l\neq 0.$ For 
$l=0,$ $\mu \neq 0,$ these states become singular at $r=0.$ However they
still can be normalized to a $\delta -$function. The states of the second
type ($j=2$) vanish at $r=0$ for any $l.$

The arbitrary constant spinor $\upsilon $ from (\ref{35}) can be specified
by an appropriate choice of spin integrals of motion \cite{BagGi90}. In what
follows we are going to write $\upsilon $ as 
\begin{equation}
\upsilon =\frac{1}{2}\left( 
\begin{array}{c}
1+\zeta \\ 
1-\zeta
\end{array}
\right) ,\;\zeta =\pm 1\,.  \label{39}
\end{equation}
In this case, $\zeta =+1$ corresponds to the spin along the magnetic field,
and $\zeta =-1\,$corresponds to the spin opposite to the magnetic field.

Finally, we briefly review classical motion of a charged particle in the
magnetic solenoid field. That is useful for an interpretation of quantum
numbers in the problem under consideration. Suppose we consider classical
trajectories that do not intersect $z$ axis. Such trajectories are not
affected by the solenoid field and have the form 
\begin{eqnarray}
&&x^{0}=\frac{k_{0}}{m}\tau ,\;x^{1}=R\cos \kappa +x_{\left( 0\right)
}^{1},\;x^{2}=R\sin \kappa +x_{\left( 0\right) }^{2}\,,\;x^{3}=-\frac{k_{3}}{%
m}\left( \tau -\tau _{0}\right) \,,  \nonumber \\
\; &&\kappa =\omega _{0}\tau +\varphi _{0}\,,\;\omega _{0}=\frac{\gamma }{m}%
\,\,,\;k_{0}^{2}=m^{2}+k_{3}^{2}+\gamma ^{2}R^{2}\,\,.  \label{41}
\end{eqnarray}
Here$\ \tau \ $is the relativistic interval and $R,\varphi _{0},x_{\left(
0\right) }^{1},x_{\left( 0\right) }^{2},\tau _{0},k_{0},k_{3}$ are
integration constants. Classical analogs of quantum operators $\hat{P}_{\mu
} $ and $\hat{L}_{z}$ read 
\begin{eqnarray}
&&P_{0}=\hbar k_{0},\;P_{1}=\hbar \gamma R\,\sin \kappa ,\;P_{2}=-\hbar
\gamma R\,\cos \kappa \,,\;P_{3}=\hbar k_{3}\,,\;  \nonumber \\
\; &&L_{z}\,=\hbar \frac{\gamma }{2}\left( R^{2}-R_{0}^{2}\right) -\hbar
\left( l_{0}+\mu \right) \,,\,\,R_{0\,}^{2}=\left( x_{\left( 0\right)
}^{1}\right) ^{2}+\left( x_{\left( 0\right) }^{2}\right) ^{2}.  \label{43}
\end{eqnarray}
On the plane $z={\rm const,}$ the trajectories (\ref{41}) are circles $%
\left( x^{1}-x_{\left( 0\right) }^{1}\right) ^{2}+\left( x^{2}-x_{\left(
0\right) }^{2}\right) ^{2}=R^{2}$ of radius $R.$ The motion along the axis $%
z $ is uniform with the velocity $v_{3}=ck_{3}/k_{0}\,.$ Comparing the
classical radial momentum $P_{r}^{2}=P_{1}^{2}+P_{2}^{2}$ with the
corresponding quantum expressions (\ref{14}), (\ref{16}), we get 
\begin{equation}
R^{2}=\frac{2\bar{n}+1}{\gamma }\,.  \label{46}
\end{equation}
This equation\ relates the principal quantum number to the radius $R$ of the
classical motion. Comparing $L_{z}$ from (\ref{43}) with the corresponding
quantum expression (\ref{10}), we find 
\begin{equation}
\bar{l}=l+\mu =\frac{\gamma }{2}\left( R^{2}-R_{0}^{2}\right) \,.  \label{47}
\end{equation}
Thus, we can conclude that classical trajectories with $l\geq -\mu $ embrace
the solenoid ($R^{2}>R_{0}^{2}$) and ones with $l<-\mu $ do not. In quantum
theory these conditions are $l\geq 0$ and $l<0$ respectively. A minimal
distance $\Delta R$ between a classical trajectory and the solenoid is
related to $\,l+\mu $ as follows 
\begin{equation}
\Delta R=\left| R-R_{0}\right| =\frac{2|l+\mu |}{\gamma \left(
R+R_{0}\right) }\,\,.  \label{48}
\end{equation}
Thus, in fact, the absolute value of $l$ specifies the above distance.

Trajectories with $l=0$ and$\;l=-1$ pass most close to the solenoid. In the
first case they embrace the solenoid and in the second one do not. As was
already mentioned above, Dirac wave functions with $l=0$ are singular at $%
r=0.$ Bearing in mind the classical interpretation of such trajectories, we
may treat the existence of the singularity as a result of a superstrong
interaction between the electron spin and the solenoid.

The wave functions (\ref{32}) with $N_{D}=[8\pi Lk_{0}(k_{0}+m)/\gamma
]^{-1/2}\;$obey the following orthonormality relations ($-L<z<L$, $%
L\rightarrow \infty $) 
\begin{equation}
(\Psi _{n^{\prime },l^{\prime },k_{3}^{\prime }},\Psi _{n,l,k_{3}})=\int
\Psi _{n^{\prime },l^{\prime },k_{3}^{\prime }}^{+}\Psi _{n,l,k_{3}}d{\bf r}%
=\delta _{n,n^{\prime }}\delta _{l,l^{\prime }}\delta _{k_{3},k_{3}^{\prime
}}(v^{\prime }{}^{+}\,v)\;.  \label{48a}
\end{equation}

\section{Matrix elements of electron transitions with one photon radiation}

In QED, one-photon radiation intensity caused by electron transitions is
given by the expression 
\begin{eqnarray}
&&W_{\lambda }=\frac{ce^{2}}{2\pi }\int d{%
\mbox{\boldmath$\kappa
$\unboldmath}}\delta \left( \kappa +k_{0}^{a}-k_{0}^{b}\right) \left| \bar{%
\mbox{\boldmath$\alpha $
\unboldmath}}{\bf e}_{\lambda }\right| ^{2},  \nonumber \\
&&{\mbox{\boldmath$\kappa $\unboldmath}}=\kappa \left( \sin \theta \cos
\varphi ^{\prime },\sin \theta \sin \varphi ^{\prime },\cos \theta \right)
\;,  \label{49}
\end{eqnarray}
where ${\mbox{\boldmath$\kappa $\unboldmath}}$\ is photon wave vector \cite
{SokTe68}. Spherical angles $\theta ,\,\varphi ^{\prime }$ define angular
distribution of the emitted photons and $\kappa =\left| {%
\mbox{\boldmath$\kappa
$\unboldmath}}\right| $ defines the frequency $\omega =c\kappa $ and the
energy $E_{ph}=c\hbar \kappa $ of a photon. The quantities $%
k_{0}^{a},k_{0}^{b}$ are related to electron energies $E^{a},E^{b}$ in
initial and final states as $E^{a,b}=chk_{0}^{a,b}$. Unit vectors ${\bf e}%
_{\lambda }$ characterize radiation polarization, see for example \cite
{SokTe68}. $\bar{{\mbox{\boldmath$\alpha
$\unboldmath}}}$ denotes a matrix element of the operator ${%
\mbox{\boldmath$\alpha$\unboldmath}}=\left( \alpha ^{i}=\gamma ^{0}\gamma
^{i}\right) ,\;$ 
\begin{equation}
\bar{{\mbox{\boldmath$\alpha $
\unboldmath}}}=\int d{\bf x}\Psi _{a}^{+}\left( x\right) e^{-i{%
\mbox{\boldmath$\kappa $\unboldmath}}{\bf x}}{%
\mbox{\boldmath$\alpha
$\unboldmath}}\Psi _{b}\left( x\right) \,.  \label{51}
\end{equation}
For spinless particle case, one has to replace ${\bf \bar{\alpha}}$ by ${\bf 
\bar{P},}$ 
\begin{equation}
{\bf \bar{P}=}\int d{\bf x}\Psi _{a}^{\ast }\left( x\right) e^{-i{%
\mbox{\boldmath$\kappa $\unboldmath}}{\bf x}}{\bf \hat{P}}\Psi _{b}\left(
x\right) \,,  \label{52}
\end{equation}
where ${\bf \hat{P}=}\left( \hat{P}^{i}\right) .$ To get total intensity of
the polarized radiation, we have to sum (\ref{49}) over all the final states
of the electron. In the SR theory \cite{SokTe68}, a linear polarization of
the radiation is described by $\sigma $ and $\pi $ components of the
operator ${\bf \hat{P},}$ 
\begin{equation}
\hat{P}_{\sigma }=\hat{P}_{1}\sin \varphi ^{\prime }-\hat{P}_{2}\cos \varphi
^{\prime },\,\hat{P}_{\pi }=-(\hat{P}_{1}\cos \varphi ^{\prime }+\hat{P}%
_{2}\sin \varphi ^{\prime })\cos \theta +\hat{P}_{3}\sin \theta .  \label{53}
\end{equation}
Using Eq. (\ref{6}), these components can be written as 
\begin{eqnarray}
&&\hat{P}_{\sigma }=i\hbar \sqrt{\frac{\gamma \rho }{2}}\left( A-B\right)
\,,\;\hat{P}_{\pi }=i\hbar \sqrt{\frac{\gamma \rho }{2}}\left( A+B\right)
+i\hbar \sin \theta \frac{\partial }{\partial x^{3}}\,,  \nonumber \\
A &=&e^{i\left( \varphi ^{\prime }-\varphi \right) }\left( \frac{\rho
+l_{0}+\mu -i\partial _{\varphi }}{2\rho }-\partial _{\rho }\right)
,\;B=e^{-i\left( \varphi ^{\prime }-\varphi \right) }\left( \frac{\rho
+l_{0}+\mu -i\partial _{\varphi }}{2\rho }+\partial _{\rho }\right)
\label{54}
\end{eqnarray}

Consider first the spinless particle case. Thus, we have to substitute (\ref
{54}) and functions (\ref{15}), (\ref{24}) into (\ref{52}). We are going to
mark off quantum numbers of final states by primes. The integration over $%
x^{3}$ in (\ref{52}) leads to a conservation low for $z-$component of the
momentum 
\begin{equation}
k_{3}-k_{3}^{\prime }=\kappa \cos \theta \;.  \label{55}
\end{equation}
Integrating over $\kappa $ in (\ref{49}), we get a conservation low for the
energy, 
\begin{equation}
k_{0}-k_{0}^{\prime }-\kappa =0\,.  \label{56}
\end{equation}
Doing integration over $\varphi ,$ we meet integrals of the form 
\begin{eqnarray}
J &=&\frac{1}{2\pi }\int_{0}^{2\pi }\exp [i(l-l^{\prime })\varphi -i\kappa
r\sin \theta \cos (\varphi -\varphi ^{\prime })]d\varphi  \nonumber \\
&=&J_{l^{\prime }-l}\left( 2\sqrt{q\rho }\right) \exp \left[ i\left(
l-l^{\prime }\right) \left( \varphi ^{\prime }+\frac{\pi }{2}\right) \right]
\,,  \label{57}
\end{eqnarray}
where 
\begin{equation}
q=\frac{\kappa ^{2}\sin ^{2}\theta }{2\gamma }\,.  \label{58}
\end{equation}
To make sure that (\ref{57}) is correct, one can use an integral
representation for Bessel functions (\cite{GraRy94}, 8.411.1). Integrating
over $\rho ,$ we meet two integrals containing the Laguerre functions. These
integrals can be done exactly as well, 
\begin{eqnarray}
&&\int\limits_{0}^{\infty }{I_{\alpha +m,m}(x)I_{\beta +n,n}(x)J_{\alpha
+\beta }(2\sqrt{qx})dx}=(-1)^{n+m}I_{\beta +n,m}(q)I_{\alpha +m,n}(q),\; 
\nonumber \\
\; &&0\leq n,m\in Z,\;\Re \,(\alpha +\beta +1)>0\,;  \label{59} \\
&&\int\limits_{0}^{\infty }{I_{\alpha +m,m}(x)I_{\beta +n,n}(x)J_{\alpha
-\beta }(2\sqrt{qx})dx}=(-1)^{n+m}I_{n,m}(q)I_{\alpha +m,\beta +n}(q),\; 
\nonumber \\
&&\;0\leq n,m\in Z,\;\Re (\alpha +1)>0\,.  \label{60}
\end{eqnarray}
Similar integrals can be encountered in (\cite{GraRy94}, 7.422.2), however
there the calculation was in error.

Spinning particle case can be analyzed in the same manner. Thus, as in the
conventional SR theory, matrix elements of electron transitions in the
magnetic-solenoid field can be calculated exactly.

\section{Analysis of radiation frequencies}

The relations (\ref{55}) and (\ref{56}) together with ones (\ref{28}) or (%
\ref{36}) define the frequency of the radiation $\kappa $ as a function of
initial and final quantum numbers and of the angle $\theta $. Due to the
axial symmetry of the problem, $\kappa $ does not depend on $\varphi
^{\prime }$. Similar to the conventional SR theory, we introduce a number $%
\nu $ of emitted harmonic as 
\begin{equation}
\nu =n-n^{\prime }\;.  \label{61}
\end{equation}
For $\mu =0,$ the frequency $\kappa $ is a function of the principle quantum
number $n$, of $\nu ,$ and of the angle $\theta $ (see \cite{SokTe68}). This
frequency does not depend on the azimuthal quantum numbers $l,\;l^{\prime }$%
. For $\mu \neq 0,$ this degeneracy is partially lifted. In such a case, the
frequency $\kappa $ depends on the type of initial and final states. Namely,
it depends on the quantum numbers $j,\;j^{\prime }$ in accordance with Eq. (%
\ref{56}). Thus $\kappa =\kappa _{jj^{\prime }}$. Introducing an effective
number $\bar{\nu}=\bar{\nu}_{jj^{\prime }}$ of emitted harmonic as 
\begin{equation}
\bar{\nu}=\bar{n}-\bar{n}^{\prime }=\nu +\mu (j^{\prime }-j)=\left\{ 
\begin{array}{l}
\nu ,\;j=j^{\prime }\; \\ 
\nu +\mu ,\;j=1,\;j^{\prime }=2,\;\bar{\nu}>0 \\ 
\nu -\mu ,\;j=2,\;j^{\prime }=1
\end{array}
\right. \;,  \label{62}
\end{equation}
one can easily get for spinless particle case 
\begin{equation}
\kappa _{jj^{\prime }}=\frac{k_{0j}}{\sin ^{2}\theta }\left( 1-\sqrt{1-\beta
_{j}^{2}\frac{2\bar{\nu}}{2\bar{n}+1}\sin ^{2}\theta }\right) \;,  \label{63}
\end{equation}
where 
\begin{equation}
\beta _{j}^{2}=1-\left( \frac{m}{k_{0j}}\right) ^{2}=1-\left( \frac{%
m_{0}c^{2}}{E_{j}}\right) ^{2}\;.  \label{65}
\end{equation}
Similar formula takes place for spinning particle case, 
\begin{equation}
\kappa _{jj^{\prime }}=\frac{k_{0j}}{\sin ^{2}\theta }\left( 1-\sqrt{1-\beta
_{j}^{2}\frac{\bar{\nu}}{\bar{n}}\sin ^{2}\theta }\right) \;.  \label{64}
\end{equation}
The expressions (\ref{63}) and (\ref{64}) are obtained for initial states
with $k_{3}=0$ (next we use the same supposition). Both expressions can be
written in the form 
\begin{equation}
\kappa _{jj^{\prime }}=\frac{2\gamma \bar{\nu}}{k_{0j}+\sqrt{%
k_{0j}^{2}-2\gamma \bar{\nu}\sin ^{2}\theta }}\;.  \label{66}
\end{equation}
We can also get the following formulas 
\begin{eqnarray}
&&\kappa =\frac{\gamma }{k_{0j}}(\bar{\nu}+q),\;\;\sin \theta =\sqrt{\frac{2%
}{\gamma }}k_{0j}\frac{\sqrt{q}}{\bar{\nu}+q}\;,  \nonumber \\
&&\sqrt{k_{0j}^{2}-2\gamma \bar{\nu}\sin ^{2}\theta }=k_{0j}\frac{\bar{\nu}-q%
}{\bar{\nu}+q}\;,  \label{67}
\end{eqnarray}
where the quantity $q$ was defined by Eq. (\ref{58}).

Thus, for $\mu \neq 0,$ there appear two spectral series: one results from
transitions without any change of the quantum number $j$ and another one
results from transitions with the change of $j$. In the former case $\bar{\nu%
}=\nu $ whereas in the latter case $\bar{\nu}=\nu \pm \mu $ (effective
numbers of emitted harmonics are not integer anymore). Whenever $\nu >0$ and 
$n$ are fixed, we gets an inequality 
\begin{equation}
\kappa _{21}<\kappa _{11}<\kappa _{22}<\kappa _{12}\;,  \label{68}
\end{equation}
which becomes the equality for $\mu =0$. The difference between the
frequencies $\kappa _{11}$ and $\kappa _{22}$ can be easily estimated for
laboratory magnetic fields $H\ll H_{0}$, where $H_{0}=m_{0}^{2}c^{3}/e\hbar
\approx 4,41\times 10^{13}{\rm gauss}$ is a critical field. This difference
is proportional to $\mu $, 
\begin{equation}
\kappa _{22}-\kappa _{11}\approx \mu \,\kappa _{22}\,\delta \,,\;\;\delta =%
\frac{\gamma }{k_{0}^{2}}=\frac{\gamma }{m^{2}}\frac{m^{2}}{k_{0}^{2}}=\frac{%
H}{H_{0}}\left( \frac{m_{0}c^{2}}{E_{1}}\right) ^{2}\;.  \label{70}
\end{equation}
One can see that $\delta <10^{-9}$ for not very high electron energies and
for typical (those which are realized in accelerators)\ magnetic fields $%
H\approx 10^{4}{\rm gauss}$ . In such a case, the frequency difference reads 
\begin{equation}
\kappa _{12}-\kappa _{21}\approx 2\kappa _{12}\frac{\mu }{\nu }=2\frac{%
\omega }{c}\mu \;,\;\omega =\frac{\left| ecH\right| }{E_{j}}\,,  \label{71}
\end{equation}
where $\omega $ is the synchrotron frequency. Thus, this difference becomes
quite noticeable for harmonics with small numbers.

For $\mu \neq 0,$ there exist a radiation of a harmonic $\nu =0$ due to $%
j=1\rightarrow $ $j^{\prime }=2$ transitions. For magnetic fields $H\ll
H_{0} $ the frequency of such a harmonic is 
\begin{equation}
\omega _{12}=\mu \omega \,=\mu \frac{ecH}{E_{1}}\,.  \label{72}
\end{equation}
$j=2\rightarrow $ $j^{\prime }=1$ transitions cause a radiation of a
harmonic $\nu =1.$ For the above magnetic fields the corresponding frequency
reads 
\begin{equation}
\omega _{21}=(1-\mu )\omega \,.  \label{73}
\end{equation}
Such a radiation does not exist in pure magnetic field. Both frequencies $%
\omega _{12},\omega _{12}$ are less than $\omega $ ($\omega $ is the least
radiated frequency in pure magnetic field).

\section{Radiation of spinless particle}

\subsection{Exact expression for radiation intensity}

As was discussed above, transition matrix elements that define one-photon
radiation in the magnetic-solenoid field can be calculated exactly. For
spinless particle case, the differential (with respect to the polarization)
radiation intensity has the form: 
\begin{equation}
W_{j}=W_{0}\frac{H}{H_{0}}\left( \frac{\gamma }{k_{0j}^{2}}\right)
^{2}\sum_{\nu ,j^{\prime }}\frac{1}{2\pi }\int_{0}^{2\pi }d\varphi ^{\prime }%
\frac{1}{2}\int_{0}^{\pi }d\theta \,\sin \theta \frac{(\bar{\nu}+q)^{3}}{%
\bar{\nu}-q}Q_{jj^{\prime }}|F_{jj^{\prime }}|^{2}\;,  \label{74}
\end{equation}
where 
\begin{eqnarray}
&&W_{0}=\frac{e^{2}m_{0}^{2}c^{3}}{\hbar ^{2}}\,,\;F_{jj^{\prime }}=2l_{2}%
\sqrt{q}I_{jj^{\prime }}^{\prime }(q)+l_{3}\cot \theta \sqrt{\frac{%
2k_{0j}^{2}}{\gamma }}I_{jj^{\prime }}(q)\;,  \nonumber \\
&&I_{1j^{\prime }}(q)=I_{\bar{n},\bar{n}^{\prime }}(q),\;\;I_{2j^{\prime
}}(q)=I_{\bar{n}^{\prime },\bar{n}}(q)\;.  \label{76}
\end{eqnarray}
The quantities $F_{jj^{\prime }}$ do not depend on the azimuthal quantum
number $l.$ These quantities are completely defined by the quantum numbers $%
\bar{n},\bar{\nu}$,$j,j^{\prime }$ and by the polarization of the radiation.
The polarization is characterized by quantities $l_{2}$ and $l_{3}$ (see 
\cite{SokTe68}). For $l_{2}=1,\;l_{3}=0,$ we get so called $\sigma $%
-component of the linear polarization; for $l_{2}=0,\;l_{3}=1$ we get so
called $\pi $-component of the linear polarization; for $l_{2}=\pm l_{3}=1/%
\sqrt{2},$ we get right (left) circular polarization, and, finally, for $%
l_{2}^{2}=l_{3}^{2}=1,\;l_{2}\cdot l_{3}=0,$ we get total intensity of
non-polarized radiation. The quantities $Q_{jj^{\prime }}$ depend on initial
quantum number $l$ only, 
\begin{equation}
Q_{1j^{\prime }}=\sum_{l^{\prime }}I_{\bar{n}^{\prime }-\bar{l}^{\prime },%
\bar{n}-\bar{l}}^{2}(q),\;\;Q_{2j^{\prime }}=\sum_{l^{\prime }}I_{\bar{n}-%
\bar{l},\bar{n}^{\prime }-\bar{l}^{\prime }}^{2}(q)\;.  \label{77}
\end{equation}
Limits of the summation over final quantum numbers $l^{\prime }$ depend on $%
j^{\prime }$. Namely, $0\leq l^{\prime }\leq n-\nu $ whenever $j^{\prime
}=1, $ and $-\infty <l^{\prime }\leq -1$ whenever $j^{\prime }=2$.

The integrand in (\ref{74}) does not depend on $\varphi ^{\prime }$. Thus,
the integration over this angle is trivial. The corresponding factor will be
taken into account in following expressions.

Integrating over $\theta ,$ we get zero for total circular polarization. The
reason is that dominant circular polarizations in the upper ($0\leq \theta
\leq \pi /2$) and in the lower ($\pi /2\leq \theta \leq \pi $) half-planes
have opposite signs and compensate each other exactly. If we are interested
in linear polarization only, then we can always set $l_{2}\cdot \,l_{3}=0$
in Eq. (\ref{74}).

Now we are going to fulfil summation in the intensity of the radiation over
final azimuthal quantum numbers .

Consider first the case $\mu =0.$ Here the quantity $|F|^{2}$ in Eq. (\ref
{74}) does not depend on the type of the final state since the property (\ref
{22}) is valid in this case. Effective quantum numbers coincide with
ordinary ones, $\bar{n}=n,\;\bar{l}=l$. Thus, taking into account (\ref{77}%
), we get 
\begin{equation}
\sum_{j^{\prime }}Q_{jj^{\prime }}=\sum_{k=0}^{\infty
}I_{k,s}^{2}(q)=1,\;s=n-l\;.  \label{80}
\end{equation}
That is a well-known result in SR theory \cite{SokTe68}. In other words, the
radiation intensity does not depend on the initial azimuthal quantum number $%
l$.

In magnetic-solenoid field with $\mu \neq 0,$ the quantities $Q_{jj^{\prime
}}$ depend on $l,\,n,\,\nu $. Thus, the degeneracy with respect to $l$ is
lifted completely. That may be interpreted as follows: According to Eq. (\ref
{48}), the quantum numbers $l$ define distances between classical
trajectories and the solenoid. At the same time, $l$ defines the type of
trajectories. Clearly, that the radiation intensity depends on the distances
as well as on the type of states. For $\mu =0,$ the origin is not fixed
anymore by the presence of the solenoid. Thus, the $l$ dependence of the
radiation intensity dies out.

Let us return to Eq.(\ref{77}), which define $Q_{jj^{\prime }}$. Using
properties of the Laguerre functions, we can get the following expression
for a derivative of $Q_{jj^{\prime }}$ 
\begin{eqnarray}
&&\frac{d}{dq}Q_{jj^{\prime }}(q)=(-1)^{1+j+j^{\prime }}\sqrt{\frac{k+1}{q}}%
\left[ (2-j)I_{k+1,s}(q)I_{k,s}(q)\right.  \nonumber \\
&&\left. +(j-1)I_{s,k+1}(q)I_{s,k}(q)\right] \,,\;\;k=\bar{n}^{\prime }-\mu
,\;s=\bar{n}-\bar{l}\;.  \label{81}
\end{eqnarray}
Then, taking into account the behavior of $Q_{jj^{\prime }}$ at $q=0$ and\
at $q=\infty $, we obtain 
\begin{eqnarray}
&&Q_{jj^{\prime }}(q)=j^{\prime }-1+(-1)^{j^{\prime }-1}\left[
(2-j)\int_{q}^{\infty }\sqrt{\frac{k+1}{y}}I_{k+1,s}(y)I_{k,s}(y)dy\right. 
\nonumber \\
&&\left. +(j-1)\int_{0}^{q}\sqrt{\frac{k+1}{y}}I_{s,k+1}(y)I_{s,k}(y)dy%
\right] \,.  \label{82}
\end{eqnarray}
The result (\ref{80}) follows from (\ref{82}) as $\mu \rightarrow 0$.

\subsection{Radiation in weak magnetic field approximation}

Consider here the magnetic-solenoid field with $H$ obeying the condition $%
H\ll H_{0}$\thinspace . Besides, we suppose that 
\begin{equation}
2\gamma \bar{\nu}k_{0j}^{-2}=2\frac{H}{H_{0}}\left( \frac{m_{0}c^{2}}{E_{j}}%
\right) ^{2}\bar{\nu}\ll 1\,.  \label{83}
\end{equation}
It is known that the only $\bar{\nu}\sim (E_{j}/m_{0}c^{2})^{3}$ harmonics
are effectively emitted in the relativistic case. For such harmonics Eq. (%
\ref{83}) results in 
\begin{equation}
2\frac{H}{H_{0}}\frac{E_{j}}{m_{0}c^{2}}\ll 1\;.  \label{84}
\end{equation}
It was demonstrated in \cite{SokTe68} that the condition (\ref{84}) implies
insignificance of quantum corrections in the relativistic case. That may be
not true in the non-relativistic approximation since the only harmonic $\bar{%
\nu}\sim 1$ is emitted effectively and the condition (\ref{83}) always holds
for $H\ll H_{0}$. Practically, the condition (\ref{84}) always holds for
real laboratory magnetic fields and electron energies. In the above
suppositions, the quantity (\ref{58}) reads 
\begin{equation}
q=\frac{1}{2}\frac{H}{H_{0}}\left( \frac{m_{0}c^{2}}{E_{j}}\right) ^{2}\bar{%
\nu}^{2}\sin ^{2}\theta \;.  \label{85}
\end{equation}
Suppose that the numbers of harmonics are not very big, then we can expect
that 
\begin{equation}
q\ll 1\;.  \label{86}
\end{equation}
For the ultra-relativistic case $\nu \sim (E_{j}/m_{0}c^{2})^{3},$ we find 
\[
q\sim \frac{H}{H_{0}}\left( \frac{E_{j}}{m_{0}c^{2}}\right) ^{4}\;. 
\]
That means that in the latter case $q$ can be not small. Thus, namely the
condition (\ref{86}) defines the non-relativistic case in the weak magnetic
field approximation. In other words, the condition (\ref{86}) corresponds to
CR in weak magnetic field approximation. Namely such a radiation is of
concern to us in this Section. Below we suppose that (\ref{86}) takes place.

In the case under consideration, we can present the radiation intensity in
the following form 
\begin{eqnarray}
&&W_{j}=W_{j}^{{\rm cl}}\bar{W}_{j}\,,\;\;W_{j}^{{\rm cl}}=\frac{2}{3}\frac{%
e^{4}H^{2}\beta _{j}^{2}(1-\beta _{j}^{2})}{m_{0}^{2}c^{3}}\;,  \label{87} \\
&&\bar{W}_{j}=\sum_{j^{\prime }}\bar{W}_{jj^{\prime }}\,,\;\;\bar{W}%
_{jj^{\prime }}=\frac{3}{4(2\bar{n}+1)}\int_{0}^{p}\sin \theta d\theta
S_{jj^{\prime }}^{2}\sum_{\nu }\bar{\nu}^{2}R_{jj^{\prime }}\;,  \label{88}
\\
&&S_{11}=S_{22}=S_{12}=l_{2}+l_{3}\cos \theta =S,\;S_{21}=l_{2}-l_{3}\cos
\theta =\bar{S}\;.  \label{89}
\end{eqnarray}
The quantity $W_{j}^{{\rm cl}}$ is the radiation intensity of a first
harmonic in the semiclassical approximation (see \cite{SokTe68}). The
radiation polarization is characterized by the factor $S_{jj^{\prime }}$. In
particular, one can see that for transitions $j=2\rightarrow j^{\prime }=1$
the sign of the radiation circular polarization is opposite to the sign of
the circular polarization for all other transitions. That observation can be
useful to identify the radiation related to $j=2\rightarrow j^{\prime }=1$
transitions. The quantities $R_{jj^{\prime }}$ can be calculated in lowest
order of $q$ using exact expressions (\ref{74}), (\ref{76}), and (\ref{82}).

1. For transitions $j=1\rightarrow j^{\prime }=1,$ we find : 
\begin{eqnarray}
&&R_{11}=\frac{\Gamma (n+\mu +1)q^{\nu -1}}{\Gamma (n+\mu +1-\nu )\Gamma
^{2}(\nu )}\;,\;\;1\leq \nu \leq l\;,  \nonumber \\
&&R_{11}=\frac{\Gamma (n-l+1)\Gamma (n+\mu +1)q^{2\nu -l-1}}{\Gamma (n-\nu
+1)\Gamma (n-\nu +\mu +1\Gamma ^{2}(\nu -l+1))\Gamma ^{2}(\nu )}\;,\;\;l\leq
\nu \leq n\;.  \label{90}
\end{eqnarray}
One can see that the only harmonic $\nu =1$ is effectively emitted in these
transitions. At the same time, the radiation intensity does not depend on $l$
whenever\ $1\leq l\leq n$. The quantity $\bar{W}_{11}$ can be easily
calculated, 
\begin{equation}
\bar{W}_{11}=\frac{2(n+\mu )}{2(n+\mu )+1}\bar{S^{2}}\,,\;\;\bar{S^{2}}=%
\frac{3}{4}l_{2}^{2}+\frac{1}{4}l_{3}^{2}\;.  \label{91}
\end{equation}
Thus, the radiation is polarized similarly to the conventional ($\mu =0$) SR
case \cite{SokTe68}. The quantity $\bar{W}_{11}$ increases as $n\rightarrow
\infty $, in particular, $\lim_{n\rightarrow \infty }\bar{W}_{11}=\bar{S^{2}}
$ .

For initial states with $l=0,$ transition probabilities are of order $q$.
Let $\nu =1$, then we find for such transitions 
\begin{equation}
\bar{W}_{11}=\frac{H}{H_{0}}\left( \frac{m_{0}c^{2}}{E_{1}}\right) ^{2}\frac{%
3n(n+\mu )}{5(2n+2\mu +1)}\left( \frac{5}{6}l_{2}^{2}+\frac{1}{6}%
l_{3}^{2}\right) \,.  \label{93}
\end{equation}
Here the linear polarization of the radiation is greater than in (\ref{91}).
However, these transitions contribute insignificantly to the radiation
compared to all other transitions.

2. For transitions $j=2\rightarrow j^{\prime }=2,$ we find 
\begin{equation}
R_{22}=\frac{\Gamma (n+1)q^{\nu -1}}{\Gamma (n+1-\nu )\Gamma ^{2}(\nu )}\;.
\label{94}
\end{equation}
As before, we see that the only first harmonic is effectively emitted. For
this harmonic $R_{22}=n$ and 
\begin{equation}
\bar{W}_{22}=\frac{2n}{2n+1}\bar{S^{2}}\;.  \label{95}
\end{equation}
(Eq. (\ref{95}) follows from (\ref{91}) as $\mu \rightarrow 0$.) The
radiation\ intensity does not depend on $l<0$ in the least order of $q$.

3. For transitions $j=2\rightarrow j^{\prime }=1,$ we find 
\begin{equation}
R_{21}=\frac{\Gamma (n+|l|+1-\mu )\Gamma (n-\nu +1+\mu )\Gamma ^{2}(1+\nu
-\mu )\mu ^{2}(1-\mu )^{2}q^{|l|-1}}{\Gamma (n+1-\nu )\Gamma (n+1)\Gamma
^{2}(|l|+\nu +1-\mu )}\,f^{2}(\mu )\;.  \label{96}
\end{equation}
We have introduced here a function $f(\mu )$, $0\leq \mu \leq 1,\;$ 
\begin{eqnarray}
&&f(\mu )=\frac{\sin \mu \pi }{\mu (1-\mu )\pi }\,,\;f(\mu )=f(1-\mu
),\;f(0)=f(1)=1,\;  \nonumber \\
&&\;f_{{\rm max}}(\mu =1/2)=4/\pi >1,\;1\leq f(\mu )\leq 4/\pi \,,
\label{97}
\end{eqnarray}
which differs insignificantly from the unit whenever $\mu \neq 0$.

In the transitions under consideration, we meet a situation, which is
completely different from the one considered before. Here the only
transitions from states with $l=-1$ really contribute to the radiation. That
fact has a natural physical explanation: For $l=-1,\;j=2,$ classical
trajectories do not embrace the solenoid but pass maximally close to the
latter. A transition to trajectories embracing the solenoid is more likely
namely from such states. It is important to stress that no restrictions
exist on numbers of emitted harmonics. For $l=-1,$ we get 
\begin{equation}
R_{21}=\frac{\Gamma (n+2-\mu )\Gamma (n-\nu +1+\mu )\mu ^{2}(1-\mu )^{2}}{%
\Gamma (n+1)\Gamma (n-\nu +1)(\nu +1-\mu )^{2}}\,f^{2}(\mu )\;,  \label{99}
\end{equation}
and 
\begin{eqnarray}
&&\bar{W}_{21}=\frac{2\Gamma (n+2-\mu )\mu ^{2}(1-\mu )^{2}M^{21}f^{2}(\mu )%
}{(2n+1)\Gamma (n+1)}\,\bar{S^{2}}\;,  \nonumber \\
\text{{}} &&M^{21}=\sum_{\nu =1}^{n}M_{\nu }^{21}\,,\;\;M_{\nu }^{21}=\frac{%
\Gamma (n-\nu +1+\mu )}{\Gamma (n-\nu +1)}\left( \frac{\nu -\mu }{\nu -\mu +1%
}\right) ^{2}\;.  \label{100}
\end{eqnarray}
For example, for $n=1$ we obtain 
\begin{equation}
\bar{W}_{21}(n=1)=\frac{2\mu ^{2}(1-\mu )^{4}}{3(2-\mu )}\,f(\mu )\bar{S^{2}}%
\;.  \label{101}
\end{equation}

Expressions for big $n$ can be calculated approximately. Let us demonstrate
how one can get an estimation for a typical sum. \bigskip The sum can be
written as 
\begin{eqnarray}
&&\sum_{\nu =0}^{n}\frac{\Gamma (n+2-\mu -\nu )}{\Gamma (n-\nu +1)}\left( 
\frac{\nu +\mu }{\nu +\mu -1}\right) ^{2}=\frac{\Gamma (n+2-\mu )\mu ^{2}}{%
\Gamma (n+1)(1-\mu )^{2}}  \nonumber \\
&&+\frac{\Gamma (n+1-\mu )n(1+\mu )^{2}}{\Gamma (n+1)\mu ^{2}}+\frac{\Gamma
(n-\mu )n(n-1)}{\Gamma (n+1)}\left( \frac{2+\mu }{1+\mu }\right) ^{2} 
\nonumber \\
&&+\sum_{\nu =3}^{n}\frac{\Gamma (n+2-\mu -\nu )}{\Gamma (n-\nu +1)}\left( 
\frac{\nu +\mu }{\nu +\mu -1}\right) ^{2}\,.  \label{102}
\end{eqnarray}
For $\nu =3$ we have an inequality 
\begin{equation}
1<\left( \frac{\nu +\mu }{\nu +\mu -1}\right) ^{2}\,<\left( \frac{3+\mu }{%
2+\mu }\right) ^{2}\,,  \label{103}
\end{equation}
which allows us to write a relation 
\begin{equation}
\sum_{\nu =3}^{n}\frac{\Gamma (n+2-\mu -\nu )}{\Gamma (n-\nu +1)}\left( 
\frac{\nu +\mu }{\nu +\mu -1}\right) ^{2}=\delta \sum_{\nu =3}^{n}\frac{%
\Gamma (n+2-\mu -\nu )}{\Gamma (n-\nu +1)}\,,  \label{104}
\end{equation}
where $\delta $ can be estimated as 
\begin{equation}
1<\delta <\left( \frac{3+\mu }{2+\mu }\right) ^{2}\,.  \label{106}
\end{equation}
The latter sum can be calculated exactly using the following well-known
relation 
\begin{equation}
\sum_{\nu =1}^{n}\frac{\Gamma (n+1+\mu -\nu )}{\Gamma (n-\nu +1)}=\frac{%
\Gamma (n+1+\mu )}{\left( 1+\mu \right) \Gamma (n)}\,.  \label{105}
\end{equation}
This estimation can be improved if we write separately four or more terms in
(\ref{102}).

In the same manner, we get the following expression for the radiation
intensity 
\begin{equation}
\bar{W}_{21}=\frac{2n}{2n+1}R_{n}\left( \mu \right) \mu ^{2}\left( 1-\mu
\right) ^{2}\left[ \left( \frac{1-\mu }{2-\mu }\right) ^{2}+\left(
n-1\right) \delta \right] \bar{S^{2}},\;\frac{1}{2}<\delta <1\,,  \label{107}
\end{equation}
where 
\begin{eqnarray}
&&R_{n}\left( \mu \right) =\frac{\Gamma (n+\mu )\Gamma (n+2-\mu )}{\Gamma
^{2}(n+1)}f^{2}\left( \mu \right) \,,\;R_{0}\left( \mu \right) =\frac{%
f\left( \mu \right) }{\mu }\,,  \nonumber \\
&&R_{1}\left( \mu \right) =\left( 2-\mu \right) f\left( \mu \right)
\,,\;R_{2}\left( \mu \right) =\frac{1}{4}\left( 1+\mu \right) \left( 2-\mu
\right) \left( 3-\mu \right) f\left( \mu \right) \,,\;...\,,  \nonumber \\
&&\frac{n+1}{n}f^{2}\left( \mu \right) \geq R_{n}\left( \mu \right) \geq
R_{n}\left( 1\right) =1\,,\;\lim_{n\rightarrow \infty }R_{n}\left( \mu
\right) =f^{2}\left( \mu \right) \,.  \label{108}
\end{eqnarray}

One can see that the quantities $M^{21}$ from (\ref{100}) change slightly as 
$\nu $ changes. Thus, at least a whole succession of first harmonics has
equal probabilities of the radiation. In this approximation, such harmonics
are not emitted for $\mu =0$ (they appear only in higher orders of $q).$ For
big $n,$ one can see that $M^{21}\approx n.$ The can serve as an additional
argument in the favor of the above observation.

The case $\nu =1$ deserves to be considered especially. As was already
remarked before, the corresponding radiation frequency (\ref{73}) is less
than the cyclotron one. It follows from (\ref{100}) that 
\begin{equation}
\bar{W}_{21}(\nu =1)=\frac{2n}{2n+1}R_{n}\left( \mu \right) \left[ \frac{\mu
(1-\mu )^{2}}{2-\mu }\right] ^{2}\bar{S^{2}}\,.  \label{109}
\end{equation}
Thus, in this case, the radiation intensity is approximately equal to the
classical one multiplied by the factor 
\begin{equation}
\left[ \frac{\mu (1-\mu )^{2}f\left( \mu \right) }{2-\mu }\right] ^{2}\,.
\label{110}
\end{equation}

4. Finally, consider transitions $j=1\rightarrow j^{\prime }=2$. In this
case we get 
\begin{equation}
R_{12}=\frac{\Gamma (n-\nu -\mu +2)\Gamma (n+\mu +1)(\nu +\mu )^{2}q^{l}}{%
\Gamma (n-l+1)\Gamma (n-\nu +1)\Gamma ^{2}(l-\nu +2-\mu )\Gamma ^{2}(\nu
+1+\mu )}\;.  \label{111}
\end{equation}
We see that the only transitions from states with $l=0$ contribute
effectively to the radiation. Classical trajectories, which pass maximally
close to the solenoid (embracing it), correspond to such initial states.
Then, a possible physical interpretation is similar to the one given above.
Thus, for $l=0$ we get 
\begin{eqnarray}
&&\bar{W}_{12}=\frac{2\Gamma (n+1+\mu )\mu ^{2}(1-\mu )^{2}M^{12}f^{2}(\mu )%
}{(2n+2\mu +1)\Gamma (n+1)}\,\bar{S^{2}}\;,  \nonumber \\
&&M^{12}=\sum_{\nu =0}^{n}M_{\nu }^{12}\,,\;\;M_{\nu }^{12}=\frac{\Gamma
(n-\nu +2-\mu )}{\Gamma (n-\nu +1)}\left( \frac{\nu +\mu }{\nu +\mu -1}%
\right) ^{2}\;.  \label{112}
\end{eqnarray}
As before, we have here a radiation of $\nu =0$ harmonic (even for $n=0$ in
the initial state). Such a radiation is forbidden for $\mu =0$. The
frequency of the corresponding radiation is given by the expression (\ref{72}%
). For $\nu =0,$ one finds 
\begin{equation}
\bar{W}_{12}(\nu =0)=\mu ^{4}\frac{2(n+\mu )}{2(n+\mu )+1}R_{n}\left( \mu
\right) \,\bar{S^{2}}\,.  \label{113}
\end{equation}
In particular, for $n=0$ we find 
\begin{equation}
\bar{W}_{12}(\nu =n=0)=\frac{2\mu ^{4}f(\mu )}{2\mu +1}\bar{S^{2}}\;.
\label{114}
\end{equation}
Thus, the radiation intensity of such a harmonic is approximately equal to
the classical intensity reduced by the factor $\mu ^{4}f(\mu )$.

In the transitions $j=1\rightarrow j^{\prime }=2$, all the harmonics with
different numbers $\nu $ contribute almost equally to the radiation
intensity since $M_{\nu }^{12}$ from (\ref{112}) does not change
significantly as $\nu $ varies. One can find the following estimation for $%
\bar{W}_{12}$ (taking into account the estimation (\ref{107}) for $\delta $) 
\begin{equation}
\bar{W}_{12}=\frac{2(n+\mu )}{2\left( n+\mu \right) +1}R_{n}\left( \mu
\right) \left[ \mu ^{4}+\frac{n\left( 1-\mu ^{2}\right) ^{2}}{n+1-\mu }+%
\frac{n\left( n-1\right) \mu ^{2}\left( 1-\mu ^{2}\right) ^{2}\delta }{%
n+1-\mu }\right] \bar{S^{2}}\;.  \label{116}
\end{equation}
The existence of the transitions under consideration may be treated as a
manifestation of the AB effect. Indeed, in the absence of the solenoid field
(more exactly for $\mu =0)$ and in the approximation under consideration,
the only $\nu =1$ harmonic survives.

\subsection{Peculiarities of radiation angular distribution}

As it is known \cite{SokTe68},\ in the relativistic case the intensity of
the conventional SR is concentrated in the vicinity of the orbit plane
within a small angular interval 
\begin{equation}
\Delta \theta \approx \frac{m_{0}c^{2}}{E}=\sqrt{1-\beta ^{2}}\,.
\label{117}
\end{equation}
In such a case, the radiation intensity is maximal for harmonics with big
numbers 
\begin{equation}
\nu \sim \left( \frac{E}{m_{0}c^{2}}\right) ^{3}  \label{118}
\end{equation}
Thus, it is widely believed that low number harmonics cannot practically be
isolated against the background of intensive high frequency radiation.

However, there exist one exclusion from this rule. Indeed, we can easily see
that in the conventional SR the intensity of all the harmonics with $\nu
\geq 2$ is exactly zero in the directions $\theta =0,\pi $ (along the
magnetic field). Besides, the radiation intensity of the first harmonic ($%
\nu =1$) is maximal along the magnetic field for any particle energy.
Moreover, the latter radiation has total circular polarization and, thus,
can be easily identified.

The presence of the solenoid field modifies both the spectrum and angular
distribution of SR. Consider, for example, the intensity of SR in the
magnetic-solenoid field in the directions $\theta =0,\pi $ and within the
infinitesimal solid angle $d\Omega =\sin \theta d\theta d\varphi ^{\prime }$
. The expressions (\ref{74}), (\ref{76}), and (\ref{77}) allow us to get the
following exact result 
\begin{equation}
4\pi \left. \frac{dW_{jj^{\prime }}}{d\Omega }\right| _{\theta =0,\pi }=W^{%
{\rm cl}}G_{jj^{\prime }}\left( l,n,\nu ;\mu \right) .  \label{119}
\end{equation}
The quantity $W^{{\rm cl}}$ is defined by Eq. (\ref{87}) and 
\begin{eqnarray}
&&G_{11}=\frac{3\left( n+\mu \right) \left( 1-\delta _{l,0}\right) \delta
_{\nu ,1}}{2n+2\mu +1}\,,\;G_{22}=\frac{3n\delta _{\nu ,1}}{2n+1}\,, 
\nonumber \\
&&G_{12}=\frac{3\left( n+\mu \right) R_{n}\left( \mu \right) \delta _{l,0}}{%
2n+2\mu +1}\left[ \mu ^{4}\delta _{\nu ,0}+\frac{n\left( 1-\mu ^{2}\right)
^{2}\delta _{\nu ,1}}{n-\mu +1}\right.  \nonumber \\
&&\left. +\frac{\mu ^{2}\left( 1-\mu \right) ^{2}\Gamma \left( n+1\right) }{%
\Gamma \left( n+2-\mu \right) }\sum_{\nu =2}^{n}\frac{\Gamma \left( n+2-\mu
-\nu \right) }{\Gamma \left( n+1-\nu \right) }\left( \frac{\nu +\mu }{\nu
+\mu -1}\right) ^{2}\right] \,,\;  \nonumber \\
&&G_{21}=\frac{3\mu ^{2}\left( 1-\mu \right) ^{2}nR_{n}\left( \mu \right)
\Gamma \left( n\right) \delta _{l,-1}}{\left( 2n+1\right) \Gamma \left(
n+\mu \right) }\sum_{\nu =1}^{n}\frac{\Gamma \left( n+1+\mu -\nu \right) }{%
\Gamma \left( n+1-\nu \right) }\left( \frac{\nu -\mu }{\nu -\mu +1}\right)
^{2}.\,  \label{120}
\end{eqnarray}
The function $R_{n}\left( \mu \right) $ is given by Eq. (\ref{108}).

Let us briefly run through some of consequences of the above expression.

Transitions without a change of the type of the initial state (without a
change of $j$) cause the only first harmonic ($\nu =1$) radiation in the
directions $\theta =0,\pi $ whenever $l\neq 0.$ This fact does not depend on
particle energies. One can see that the quantities $G_{jj}$ grow slightly
and tend to finite constant values as $n\rightarrow \infty .$ Transitions
from initial states with $l=0$ without a change of $j$ do not cause any
radiation in $\theta =0,\pi $ directions.

Transitions with a change of the type of the initial state (with a change of 
$j$) cause a radiation in the directions $\theta =0,\pi $ solely for $%
l=0,-1\;$(the solenoid is situated maximally close to a classical trajectory)%
$.$ In such cases all possible harmonics ($0\leq \nu \leq n$) are emitted
with approximately equal intensities since the quantities $G_{12},$ $G_{21}$
grow proportionally to $n.$ For $\mu =0,$ the only first harmonic radiation
survives$.$

Expressions (\ref{120}) allow us to conclude that all the transitions cause
totally circular polarized radiation in the directions $\theta =0,\pi .$
Moreover, as it follows from (\ref{89}), the sign of the circular
polarization for $j=2\rightarrow j^{\prime }=1$ transitions is opposite to
the one for all other transitions.

We believe that the peculiarities of the angular distribution of the
radiation open up possibilities for experimental observation of superlow
frequencies (\ref{72}), (\ref{73})\ and of frequencies that are not multiple
of the synchrotron one.

Note, that the expressions (\ref{120}) (and the above mentioned consequences
from them) were not known before even in the absence of the solenoid field
(for $\mu =0$).

\subsection{Semiclassical approximation}

It was shown in the conventional SR theory \cite{SokTe68} that a
semiclassical expansion of the radiation intensity can be done in terms of a
small parameter $\nu /n$. Practically, to this end the formula 
\begin{equation}
{\lim_{p\rightarrow \infty }}{I_{p+\alpha ,p+\beta }\left( {\frac{x^{2}}{4p}}%
\right) }=J_{\alpha -\beta }(x)\;  \label{121}
\end{equation}
was used. Here $J_{\alpha }(x)$ are Bessel functions. It is natural to
believe that for the case $\mu \neq 0$ we can use the same parameter to
perform the semiclassical expansion. Thus, we get a classical part of the
intensity 
\begin{eqnarray}
&&W_{j}^{{\rm cl}}=\frac{e^{4}H^{2}(1-\beta _{j}^{2})}{m_{0}^{2}c^{3}}%
\sum_{\nu ,j^{\prime }}\int_{0}^{p}\sin \theta \,d\theta \bar{\nu}%
^{2}Q_{jj^{\prime }}^{{\rm cl}}|F^{{\rm cl}}|^{2},\;F^{{\rm cl}}=l_{2}\beta
_{j}J_{jj^{\prime }}^{\prime }(z)+l_{3}\cot \theta J_{jj^{\prime }}(z)\,, 
\nonumber \\
\; &&z=\bar{\nu}\beta _{j}\sin \theta \,,\;J_{11}=J_{22}=J_{12}=J_{\bar{\nu}%
}(z)\,,\;\;J_{21}=J_{-\bar{\nu}}(z)\;,  \nonumber \\
&&Q_{jj^{\prime }}^{{\rm cl}}=\frac{1}{2}+(-1)^{j+j^{\prime
}}\int_{z}^{\infty }dy(2-j)J_{l-\bar{\nu}+1}(y)J_{l-\bar{\nu}%
}(y)+(j-1)J_{|l|+\bar{\nu}}(y)J_{|l|+\bar{\nu}-1}(y)\;.  \label{122}
\end{eqnarray}
The components of $Q_{jj^{\prime }}^{{\rm cl}}$ have the form 
\begin{eqnarray}
&&Q_{11}^{{\rm cl}}=\left\{ 
\begin{array}{l}
1-\int_{0}^{z}J_{l-\nu +1}(y)J_{l-\nu }(y)dy,\;l\geq \nu \; \\ 
\int_{0}^{z}J_{\nu -l-1}(y)J_{\nu -l}(y)dy,\;l<\nu ,\;
\end{array}
\right. \;Q_{22}^{{\rm cl}}=1-\int_{0}^{z}J_{|l|+\nu }(y)J_{|l|+\nu -l}(y)dy,
\nonumber \\
&&Q_{12}^{{\rm cl}}=\left\{ 
\begin{array}{l}
\int_{0}^{z}J_{l-\nu +1-\mu }(y)J_{l-\nu -\mu }(y)dy,\;l\geq \nu \; \\ 
\frac{1}{2}-\int_{z}^{\infty }J_{l-\nu +1-\mu }(y)J_{l-\nu -\mu
}(y)dy,\;l<\nu ,\;
\end{array}
\right. \;Q_{21}^{{\rm cl}}=\int_{0}^{z}J_{|l|+\nu -\mu }(y)J_{|l|+\nu
-l-\mu }(y)dy.  \label{127}
\end{eqnarray}
For $\mu =0,$ we see that $|F^{{\rm cl}}|^{2}$ does not depend on $j^{\prime
}$ and $\sum_{j^{\prime }}Q_{jj^{\prime }}^{{\rm cl}}=1.$ Then the
expression (\ref{122}) presents the well-known \cite{SokTe68} classical SR
differential intensity.

In the non-relativistic approximation, we get 
\begin{equation}
W_{j}^{{\rm cl}}=W^{{\rm cl}}\bar{S^{2}}\sum_{j^{\prime }}\bar{W}%
_{jj^{\prime }}^{{\rm cl}}\;,\;\bar{W}_{jj}^{{\rm cl}}=1,\;\bar{W}_{12}^{%
{\rm cl}}=\mu ^{4}f^{2}(\mu ),\;\bar{W}_{21}^{{\rm cl}}=\left[ \frac{\mu
(1-\mu )^{2}f(\mu )}{2-\mu }\right] ^{2}\;.  \label{131}
\end{equation}
Here $l=\nu =0$ for $j=1\rightarrow j^{\prime }=2$ transitions, and $|l|=\nu
=1$ for $j=2\rightarrow j^{\prime }=1$ transitions. Eqs. (\ref{131}) follow
from the exact quantum expressions (\ref{95}), (\ref{110}), and (\ref{113})
as $n\rightarrow \infty $.

Semiclassical expressions (\ref{127}) depend essentially on the initial
azimuthal quantum number $l$ whenever $\mu \neq 0$. Taking into account Eq. (%
\ref{47}), we can express $l$ in terms of the pure classical quantity $%
R^{2}-R_{0}^{2}.$ If $R^{2}-R_{0}^{2}$ is fixed, we get $|l|\sim 1/\hbar
\rightarrow \infty $ as $\hbar \rightarrow 0$. Then, it follows from (\ref
{127}) that 
\begin{equation}
Q_{jj^{\prime }}^{{\rm cl}}=\delta _{jj^{\prime }}\;.  \label{133}
\end{equation}
Such a result seams to be natural. In classical theory of radiation
trajectories of particles are fixed (there is no back reaction from emitted
photons) and transitions with a change of initial states are not considered.

From Eqs. (\ref{127}), we find the following relations 
\begin{equation}
Q_{11}^{{\rm cl}}=\sum_{k=-\infty }^{l-\nu }J_{k}^{2}(z)\,,\;Q_{12}^{{\rm cl}%
}=\sum_{k=l-\nu +1}^{\infty }J_{k-\mu }^{2}(z)\,,\;Q_{22}^{{\rm cl}%
}=\sum_{k=-\infty }^{|l|+\nu -1}J_{k}^{2}(z)\,,\;Q_{21}^{{\rm cl}%
}=\sum_{k=|l|+\nu }^{\infty }J_{k-\mu }^{2}(z)\,\,.  \label{134}
\end{equation}
It is clear that $Q_{jj}^{{\rm cl}}$ are monotonically increasing functions
and $Q_{jj^{\prime }}^{{\rm cl}}$ ($j\neq j^{\prime }$) are monotonically
decreasing functions of $\left| l\right| $ such that 
\begin{equation}
\lim_{\left| l\right| \rightarrow \infty }Q_{jj^{\prime }}^{{\rm cl}}=\delta
_{_{jj^{\prime }}}\,.  \label{135}
\end{equation}
Thus, manifestations of the AB effect in SR are maximal for initial states
with $l=0,-1.$

All the angular distribution peculiarities, which were noted for the general
case in the previous Section, take place in the approximation under
consideration as well. Calculating the quantity $4\pi \left. \frac{%
dW_{jj^{\prime }}}{d\Omega }\right| _{\theta =0,\pi }$ by the help of Eqs. (%
\ref{122}) and (\ref{134}), we can see that the representation (\ref{119})
holds provided 
\begin{eqnarray}
&&G_{11}=\frac{3}{2}\left( 1-\delta _{l,0}\right) \delta _{\nu
,1}\,,\;G_{22}=\frac{3}{2}\delta _{\nu ,1}\,,\;G_{12}=\frac{3}{2}f^{2}\left(
\mu \right) \delta _{l,0}\left[ \mu ^{4}\delta _{\nu ,0}+\left( 1-\mu
^{2}\right) ^{2}\delta _{\nu ,1}\right.  \nonumber \\
&&\left. +\mu ^{2}\left( 1-\mu \right) ^{2}\sum_{\nu =2}^{n}\left( \frac{\nu
+\mu }{\nu +\mu -1}\right) ^{2}\right] \,,\;G_{21}=\frac{3}{2}f^{2}\left(
\mu \right) \mu ^{2}\left( 1-\mu \right) ^{2}\delta _{l,-1}\sum_{\nu
=1}^{n}\left( \frac{\nu -\mu }{\nu -\mu +1}\right) ^{2}\,.  \label{136}
\end{eqnarray}
The latter quantities can be obtained from (\ref{120}) in the limit $%
n\rightarrow \infty .$

\section{Radiation of spinning particle}

\subsection{Exact expression for radiation intensity}

The analysis of radiation frequencies presented in Sect.IV was done for
spinless particle case. However, all the qualitative results of this
analysis remain valid for the spinning particle case. That can be seen from
the corresponding exact formula (\ref{64}).

In the spinning particle case, we get the following exact expression for the
differential radiation intensity: 
\begin{equation}
W_{j}=W_{0}\left( \frac{H}{H_{0}}\right) ^{2}\varepsilon _{j}\int_{0}^{\pi
}d\theta \,\sin \theta \sum_{\nu ,j^{\prime },\zeta ^{\prime }}Q_{jj^{\prime
}}\left[ 1-2p_{j}\left( \bar{\nu}+q\right) \right] ^{-1}\frac{(\bar{\nu}%
+q)^{3}}{\bar{\nu}-q}|F_{jj^{\prime }}|^{2}\;.  \label{137}
\end{equation}
Here 
\begin{equation}
\varepsilon _{j}=1-\beta _{j}^{2}=\left( \frac{m}{k_{0}}\right) ^{2}=\left( 
\frac{m_{0}c^{2}}{E_{j}}\right) ^{2}\,,\;p_{j}=\frac{1}{2}\frac{H}{H_{0}}%
\frac{\varepsilon _{j}}{1+\sqrt{\varepsilon _{j}}}\,,  \label{138}
\end{equation}
and the constant $W_{0}$ was introduced in (\ref{76}). Remarkable that the
quantities $Q_{jj^{\prime }}$ are given by the same expressions (\ref{77})
as in spinless particle case. Thus, all the previous conclusions related to
these quantities are applied here as well.

In the spinning particle case, the quantities $F_{jj^{\prime }}$ have the
form 
\begin{eqnarray}
&&F_{jj^{\prime }}=l_{2}F_{jj^{\prime }}^{\left( 2\right)
}+l_{3}F_{jj^{\prime }}^{\left( 3\right) },\;\;F_{jj^{\prime }}^{\left(
2\right) }=\sqrt{\frac{H}{H_{0}}\frac{\varepsilon _{j}}{2q}}\left[ \delta
_{\zeta ,\zeta ^{\prime }}\left( -1\right) ^{j-1}\left( \frac{1+\zeta }{2}%
A_{+}^{j}+\frac{1-\zeta }{2}A_{-}^{j}\right) \right.  \nonumber \\
&&\left. +\delta _{\zeta ,-\zeta ^{\prime }}q\cot \theta \left( \frac{%
1+\zeta }{2}\chi _{+}^{j}+\frac{1-\zeta }{2}\chi _{-}^{j}\right) \right]
\,,\;F_{jj^{\prime }}^{\left( 3\right) }=\delta _{\zeta ,\zeta ^{\prime
}}\left( -1\right) ^{j-1}\cot \theta  \nonumber \\
&&\times \left( \frac{1+\zeta }{2}B_{+}^{j}+\frac{1-\zeta }{2}%
B_{-}^{j}\right) +\delta _{\zeta ,-\zeta ^{\prime }}p_{j}\left[ \left(
-1\right) ^{j-1}\frac{1+\zeta }{2}C_{+}^{j}+\frac{1-\zeta }{2}C_{-}^{j}%
\right] \,,  \label{139}
\end{eqnarray}
where 
\begin{eqnarray*}
A_{\pm }^{j} &=&2q\left[ 1-p_{j}\left( \bar{\nu}+q\right) \right] \frac{%
d\varphi _{\pm }^{j}}{dq}\mp \left[ q-p_{j}\left( \bar{\nu}+q\right) ^{2}%
\right] \varphi _{\pm }^{j}\,, \\
B_{\pm }^{j} &=&\left[ 1-p_{j}\left( \bar{\nu}+q\right) \right] \varphi
_{\pm }^{j}+2qp_{j}\,\frac{d\varphi _{\pm }^{j}}{dq},\;C_{\pm }^{j}=\left[
1\pm \sqrt{\varepsilon _{j}}\left( \bar{\nu}+q\right) \right] \chi _{\pm
}^{j}+2q\frac{d\chi _{\pm }^{j}}{dq}, \\
\varphi _{+}^{1} &=&I_{\bar{n}-1,\bar{n}^{\prime }-1}(q),\;\varphi
_{+}^{2}=I_{\bar{n}^{\prime }-1,\bar{n}-1}(q),\;\varphi _{-}^{1}=I_{\bar{n},%
\bar{n}^{\prime }}(q),\;\varphi _{-}^{2}=I_{\bar{n}^{\prime },\bar{n}}(q), \\
\chi _{+}^{1} &=&I_{\bar{n}-1,\bar{n}^{\prime }}(q),\;\chi _{+}^{2}=I_{\bar{n%
}^{\prime },\bar{n}-1}(q),\;\chi _{-}^{1}=I_{\bar{n},\bar{n}^{\prime
}-1}(q),\;\chi _{-}^{2}=I_{\bar{n}^{\prime }-1,\bar{n}}(q)\,,
\end{eqnarray*}
and $I_{n,n^{\prime }}(x)$ are the Laguerre functions. All the final quantum
numbers are primed here.The quantities $l_{2}$ and $l_{3}$ characterize the
radiation polarization. Contributions from transitions without ($\sim \delta
_{\zeta ,\zeta ^{\prime }}$) and with ($\sim \delta _{\zeta ,-\zeta ^{\prime
}}$) spin-flip are separated.

The states with $n=0$ are a special case. For $j=2,$ there exist the only
one (opposite to the magnetic field) spin orientation. Thus, all the
transitions from any states with $\zeta =-1$ to $n=0$,\thinspace\ $j=2$
states do not cause a spin-flip ($A_{+}^{j}=B_{+}^{j}=\varphi _{+}^{j}=0$),
and all the transitions from any states with $\zeta =1$ to $n=0$,\thinspace\ 
$j=2$ states do cause a spin-flip ($C_{-}^{j}=\chi _{-}^{j}=0$). One of such
transitions is studied below. States with $n=0$,\thinspace\ $j=1$ are
singular at $r=0.$ However, they still can be normalized to a $\delta -$%
function.

As in the spinless particle case, we can conclude that the radiation
intensity depends on the mantissa $\mu $ only but not on the total solenoid
magnetic flux $\Phi .$

The radiation in question has not a preferential circular polarization.
Total (integrated over all angles) intensities of the right and left
circular polarizations are equal as in the scalar particle case. However,
for transitions $j=2\rightarrow j^{\prime }=1$ and $j=2,\zeta =1\rightarrow
j^{\prime }=2,\zeta =-1$, the sign of the circular polarization is opposite
to the one for all other transitions. Further, we are going to analyze the
linear polarization only.

The angular distribution of the electron radiation intensity in the
magnetic-solenoid field is quite similar to the one for spinless particle.
The corresponding exact formula reads 
\begin{equation}
4\pi \left. \frac{dW_{jj^{\prime }}}{d\Omega }\right| _{\theta =0,\pi }=W^{%
{\rm cl}}\frac{3}{2}\left( \frac{1+\zeta }{2}G_{jj^{\prime }}^{+}+\frac{%
1-\zeta }{2}G_{jj^{\prime }}^{-}\right) ,  \label{156}
\end{equation}
where

\begin{eqnarray}
&&G_{11}^{+}=\frac{\left( n+\mu -1\right) \left( 1-\delta _{l,0}\right)
\delta _{\nu ,1}}{n+\mu }\,,\;G_{11}^{-}=\left( 1-\delta _{l,0}\right)
\delta _{\nu ,1}\,,\;G_{22}^{+}=\frac{n-1}{n}\delta _{\nu
,1}\,,\;G_{22}^{-}=\delta _{\nu ,1}\,,  \nonumber \\
&&G_{12}^{+}=\frac{nR_{n}\left( \mu \right) \delta _{l,0}}{n+\mu }\left[ \mu
^{4}\delta _{\nu ,0}+\frac{(n-1)\left( 1-\mu ^{2}\right) ^{2}\delta _{\nu ,1}%
}{n-\mu +1}+\frac{\mu ^{2}\left( 1-\mu \right) ^{2}\Gamma \left( n\right) }{%
\Gamma \left( n+2-\mu \right) }\right.  \nonumber \\
&&\left. \times \sum_{\nu =2}^{n}\frac{\Gamma \left( n+2-\mu -\nu \right) }{%
\Gamma \left( n-\nu \right) }\left( \frac{\nu +\mu }{\nu +\mu -1}\right) ^{2}%
\right] \,,\;G_{12}^{-}=R_{n}\left( \mu \right) \delta _{l,0}\left[ \mu
^{4}\delta _{\nu ,0}+\frac{n\left( 1-\mu ^{2}\right) ^{2}\delta _{\nu ,1}}{%
n-\mu +1}\right.  \nonumber \\
&&\left. +\frac{\mu ^{2}\left( 1-\mu \right) ^{2}\Gamma \left( n+1\right) }{%
\Gamma \left( n+2-\mu \right) }\sum_{\nu =2}^{n}\frac{\Gamma \left( n+2-\mu
-\nu \right) }{\Gamma \left( n+1-\nu \right) }\left( \frac{\nu +\mu }{\nu
+\mu -1}\right) ^{2}\right] \,,  \nonumber \\
&&G_{21}^{+}=\frac{R_{n}\left( \mu \right) \mu ^{2}\left( 1-\mu \right)
^{2}\Gamma \left( n+1\right) \delta _{l,-1}}{\Gamma \left( n+\mu \right) }%
\sum_{\nu =1}^{n}\frac{\Gamma \left( n+\mu -\nu \right) }{\Gamma \left(
n+1-\nu \right) }\left( \frac{\nu -\mu }{\nu -\mu +1}\right) ^{2},  \nonumber
\\
&&G_{21}^{-}=\frac{R_{n}\left( \mu \right) \mu ^{2}\left( 1-\mu \right)
^{2}\Gamma \left( n\right) \delta _{l,-1}}{\Gamma \left( n+\mu \right) }%
\sum_{\nu =1}^{n}\frac{\Gamma \left( n+1+\mu -\nu \right) }{\Gamma \left(
n+1-\nu \right) }\left( \frac{\nu -\mu }{\nu -\mu +1}\right) ^{2}.
\label{157}
\end{eqnarray}

\subsection{Radiation in weak magnetic field approximation}

Here we suppose that the magnetic field is weak, i.e. $H\ll H_{0}$ (more
exactly $q\ll 1),$ and that initial quantum numbers are not very big (thus
the particle remains non-relativistic). In this approximation, the main
contributions to the radiation are due to transitions without a spin-flip.
Below we present the only first terms in the $H/H_{0}$ decomposition for the
radiation intensity.

First consider the radiation caused by transitions without a change of the
state type ($j=j^{\prime }$). Here the transitions $\nu =1,\,l^{\prime
}=l-1\;$play the main role$.$ For initial states with $l\neq 0$ the
radiation intensity reads 
\begin{eqnarray}
&&W_{j}=W_{j}^{{\rm cl}}\left\{ \delta _{\zeta ,\zeta ^{\prime }}\left( 
\frac{1+\zeta }{2}\frac{\bar{n}-1}{\bar{n}}+\frac{1+\zeta }{2}\right)
S_{0}\right. \;  \nonumber \\
&&\left. +\delta _{\zeta ,-\zeta ^{\prime }}\left[ \frac{1+\zeta }{2}\frac{H%
}{H_{0}}\frac{S_{1}}{2\bar{n}}+\frac{1-\zeta }{2}\left( \frac{H}{H_{0}}%
\right) ^{3}\frac{\bar{n}-1}{70}S_{2}\right] \right\} \;.  \label{140}
\end{eqnarray}
The quantity $W_{j}^{{\rm cl}}$ is the radiation intensity of the first
harmonic in the semiclassical approximation (see (\ref{87})). The linear
radiation polarization is characterized by the factors 
\begin{equation}
S_{0}=\frac{3}{4}l_{2}^{2}+\frac{1}{4}l_{3}^{2}\,,\;S_{1}=\frac{1}{4}%
l_{2}^{2}+\frac{3}{4}l_{3}^{2}\,,\;S_{2}=\frac{1}{8}l_{2}^{2}+\frac{7}{8}%
l_{3}^{2}\,.  \label{141}
\end{equation}
Whenever the initial state has the spin along the field ($\zeta =1),$ the
ratio between the transitions with and without a spin-flip is of the order $%
H/H_{0}.\;$The same ratio is of the order$\;\left( H/H_{0}\right) ^{3}$ for $%
\zeta =-1.$ Thus, states with $\zeta =-1$ are more stable than ones with $%
\zeta =1.$\ That is the reason of the self-polarization effect \cite{SokTe68}
in SR. The presence of the solenoid affects the only effective quantum
numbers $\bar{n},$ the latter are not always integer, for example, $\bar{n}%
=n+\mu $ for $j=1.$ The radiation has a preferential linear polarization.
For $\zeta =1$ initial states, transitions with and without a spin-flip
cause radiation intensities of almost (with the interchange of $l_{2}$ and $%
l_{3})$ the same form $.$ For $\zeta =-1$ initial states, transitions with a
spin-flip cause almost (with the interchange of $\sigma $ and $\pi $
components) the same linear polarization of the radiation intensity as SR
has in the relativistic case \cite{SokTe68}. For $\mu =0,$ the expression (%
\ref{140}) (without the polarization specification) coincides with a
corresponding expression presented in \cite{BagDo66,TerBaD68}.

For $l=0$ in initial states, transitions without any change of $j$ are
suppressed; these transitions contribute to the radiation intensity in
higher orders of $H/H_{0}$ only. For example, for such transitions with a
spin-flip, we find 
\begin{equation}
W=W^{{\rm cl}}\frac{H}{H_{0}}\frac{3n\left( 1-\beta ^{2}\right) }{10}\left( 
\frac{1+\zeta }{2}\frac{n+\mu -1}{n+\mu }+\frac{1-\zeta }{2}\right) \left( 
\frac{5}{6}l_{2}^{2}+\frac{1}{6}l_{3}^{2}\right) \;.  \label{142}
\end{equation}
We see that for the latter transitions, the polarization is distinctive and
may serve to isolate such transitions.

Transitions with a change of the state type (with a change of $j$) are of
particular interest from the AB effect point of view. Leading (with respect
to $H/H_{0})$ contributions correspond to $j=1,l=0\rightarrow j^{\prime
}=2,l^{\prime }=-1$ and to $j=2,l=-1\rightarrow j^{\prime }=1,l^{\prime }=0$
transitions. In such transitions a whole set of successive harmonics is
emitted, all these harmonics have approximately equal probabilities. The
same situation was discovered by us in the spinless particle case.

As an example, we considered the radiation intensity for $j=1\rightarrow
j^{\prime }=2$ transitions without a spin-flip in detail. In such a case,
this intensity has the form 
\begin{equation}
W=W^{{\rm cl}}S_{0}M_{12}\delta _{\zeta ,\zeta ^{\prime }}\;,  \label{143}
\end{equation}
where $S_{0}$ is defined by (\ref{141}) and the quantity $M_{12}$ is a
function of initial quantum numbers $n$ and $\zeta ,$ 
\begin{eqnarray}
\; &&\;M_{12}=\mu ^{2}(1-\mu )^{2}f^{2}(\mu )\frac{\Gamma (n+\mu )}{\Gamma
(n+1)}  \nonumber \\
&&\times \sum_{\nu =0}^{n}\frac{\Gamma (n+2-\mu -\nu )}{\Gamma (n+1-\nu )}%
\left( \frac{\mu +\nu }{\mu +\nu -1}\right) ^{2}\left( \frac{1+\zeta }{2}%
\frac{n-\nu }{n+\mu }+\frac{1-\zeta }{2}\right) \;.  \label{144}
\end{eqnarray}
( $f(\mu )$ was defined by (\ref{97}).) In particular, here there is a
possibility for $\nu =0$ transition with the emission of a superlow
frequency (\ref{72}). For the latter transition, 
\begin{equation}
M_{12}(\nu =0)=\mu ^{4}R_{n}(\mu )\left( \frac{1+\zeta }{2}\frac{n}{n+\mu }+%
\frac{1-\zeta }{2}\right) ,  \label{145}
\end{equation}
where $R_{n}\left( \mu \right) $ is given by (\ref{108}). Similar transition
is possible even from $n=0$ states$.$ Then 
\begin{equation}
M_{12}(n=0,\zeta ,\mu )=\frac{1-\zeta }{2}\mu ^{3}f(\mu )\,.  \label{146}
\end{equation}
Taking into account that $\beta ^{2}=2\mu H/H_{0}$ for a state $j=1,\,n=0,$
we find from (\ref{143}) 
\begin{equation}
W(n=0)=\frac{4}{3}\frac{1-\zeta }{2}W_{0}\left( \frac{H}{H_{0}}\right)
^{3}\mu ^{4}\,f(\mu )S_{0}\;.  \label{148}
\end{equation}
This results fitted well with Eq. (\ref{114}) since in the spinless particle
case 
\begin{equation}
\beta ^{2}=\left( 1+2\mu \right) H/H_{0}\;  \label{149}
\end{equation}
for the state under consideration.

As before, for big $n,$ one can easily obtain estimations for emerged sums.
Thus, for the radiation intensity caused by transitions with a change of $j$
and without any spin-flip, we get the following expression : 
\begin{equation}
W=W^{{\rm cl}}R_{n}(\mu )\left( \frac{1+\zeta }{2}M^{+}+\frac{1-\zeta }{2}%
M^{-}\right) S_{0}\,,  \label{150}
\end{equation}
where 
\begin{eqnarray*}
&&M_{12}^{+}=\frac{n}{n+\mu }\left[ \mu ^{4}+\frac{\left( n-1\right) \left(
1-\mu ^{2}\right) ^{2}}{n+1-\mu }+\frac{\left( n-1\right) \left( n-2\right)
\mu ^{2}\left( 1-\mu \right) ^{2}}{n+1-\mu }\delta _{1}\right] \,, \\
&&M_{12}^{-}=\mu ^{4}+\frac{n\left( 1-\mu ^{2}\right) ^{2}}{n+1-\mu }+\frac{%
n\left( n-1\right) \mu ^{2}\left( 1-\mu \right) ^{2}}{n+1-\mu }\delta
_{1}\,,\;M_{21}^{+}=\frac{\mu \left( 1-\mu \right) ^{2}n}{n+\mu -1} \\
&&\times \left[ \mu \left( \frac{1-\mu }{2-\mu }\right) ^{2}+\left(
n-1\right) \delta _{2}\right] \,,\;M_{21}^{-}=\mu ^{2}\left( 1-\mu \right)
^{2}\left[ \left( \frac{1-\mu }{2-\mu }\right) ^{2}+\left( n-1\right) \delta
_{2}\right] \,.
\end{eqnarray*}
For $\delta _{k},\;k=1,2,$ we have the estimation$\;1<\delta
_{1}<2\,,\;1/2<\delta _{2}<1\,.$ The quantities $M_{jj^{\prime }}^{\pm }$
are linearly increasing functions of $n$ whenever $\mu \neq 0$ (the same
property takes places in spinless particle case). For $\mu =0,$ we get
another behavior 
\begin{equation}
\lim_{n\rightarrow \infty }M_{12}^{\pm }(\mu =0)=\delta _{\nu
,1},\;M_{21}^{\pm }(\mu =0,1)=0\,.  \label{152}
\end{equation}
Consider transitions that cause non zero contributions to $M_{21}^{\pm }$
for $\mu \neq 0.$ We know that the contribution of such transitions to the
radiation intensity is of higher order of $H/H_{0}$ whenever $\mu =0$. Thus,
only in the presence of the solenoid (with $\mu \neq 0)$ a whole set of
successive harmonics is emitted with approximately equal probabilities. The
number of harmonics in the set is comparable with the number of the energy
level.

For the radiation intensity caused by transitions with a change of the state
type and with the spin-flip, we get the following results:

For $j=1\rightarrow j^{\prime }=2$ transitions 
\begin{eqnarray}
&&W_{12}=W^{{\rm cl}}R_{n}(\mu )\left[ \frac{1+\zeta }{2}\frac{H}{H_{0}}%
\frac{S_{1}}{2\left( n+\mu \right) }N_{12}^{+}+\frac{1-\zeta }{2}\left( 
\frac{H}{H_{0}}\right) ^{3}\frac{2n}{35}N_{12}^{-}\right] \,,  \nonumber \\
&&N_{12}^{+}=\mu ^{6}+\frac{n(1+\mu )^{2}(1-\mu ^{2})^{2}}{n+1-\mu }\,+\frac{%
n\left( n-1\right) (n+2)^{2}\mu ^{2}(1-\mu )^{2}}{n+1-\mu }\delta _{1}\;, 
\nonumber \\
&&N_{12}^{-}=\frac{\mu ^{6}}{\left( 1+\mu \right) ^{2}}\left( \frac{\mu ^{2}%
}{8}l_{2}^{2}+\frac{7}{8}l_{3}^{2}\right) +\frac{\left( n-1\right) (1-\mu
^{2})^{2}(1+\mu )^{2}}{\left( n+1-\mu \right) \left( 2+\mu \right) ^{2}}%
\left[ \frac{(1+\mu )^{2}}{8}l_{2}^{2}+\frac{7}{8}l_{3}^{2}\right]  \nonumber
\\
&&+\frac{\left( n-1\right) \left( n-2\right) \mu ^{2}\left( 1-\mu \right)
^{2}\delta _{1}}{n+1-\mu }\left[ \frac{(n+2)^{2}}{8}l_{2}^{2}+\frac{7}{8}%
l_{3}^{2}\right] \,.  \label{153}
\end{eqnarray}

For $j=2\rightarrow j^{\prime }=1$ transitions 
\begin{eqnarray}
&&W_{21}=W^{{\rm cl}}R_{n}(\mu )\left[ \frac{1+\zeta }{2}\left( \frac{H}{%
H_{0}}\right) ^{3}\frac{2n}{35}N_{21}^{+}+\frac{1-\zeta }{2}\frac{H}{H_{0}}%
\frac{\mu (1-\mu )^{2}S_{1}}{2(n+\mu -1)}N_{21}^{-}\right] \;,  \nonumber \\
&&N_{21}^{+}=\frac{(1-\mu )^{6}}{\left( 2-\mu \right) ^{2}}\left[ \frac{%
(1-\mu )^{2}}{8}l_{2}^{2}+\frac{7}{8}l_{3}^{2}\right] +\frac{\left(
n-1\right) \mu ^{2}(2-\mu )^{4}}{\left( n+1-\mu \right) \left( 3-\mu \right)
^{2}}\left[ \frac{(2-\mu )^{2}}{8}l_{2}^{2}+\frac{7}{8}l_{3}^{2}\right] 
\nonumber \\
&&+\frac{\left( n-1\right) \left( n-2\right) \mu ^{2}\left( 1-\mu \right)
^{2}\delta _{1}}{n+1-\mu }\left[ \frac{(n+2)^{2}}{8}l_{2}^{2}+\frac{7}{8}%
l_{3}^{2}\right] \,,  \nonumber \\
&&N_{21}^{-}=\frac{\mu (1-\mu )^{4}}{\left( 2-\mu \right) ^{2}}+\left(
n-1\right) (n+2)^{2}\delta _{2\;.}  \label{154}
\end{eqnarray}
Here the radiation intensity grows as $n^{4}$ whenever $\mu \neq 0$, and the
radiation polarization depends essentially on $\mu $ and $n.$

Of special note is the loss of spin $\zeta =-1$ stability in transitions $%
j=2\rightarrow j^{\prime }=1$. It follows from (\ref{154}), for 
\begin{equation}
\delta _{3}<\mu <1-\delta _{3}\,,\;\delta _{3}=\frac{n}{3}\frac{H}{H_{0}}\,,
\label{155}
\end{equation}
that the spin $\zeta =1$ is more stable in the transitions under
considerations. Let an initial state be of second ($j=2)$ type and the
condition (\ref{155}) holds, then the radiation creates a two-phase system
of final electron states. Final electron states of second type have in the
main negative spin orientation and final electron states of first type have
in the main positive spin orientation. Thus, the presence of the solenoid
field with $\mu \neq 0$ plays a role of a depolarization factor in the above
mentioned self-polarization effect.

\subsection{Semiclassical approximation}

Consider here the radiation intensity in the semiclassical approximation.
From the previous discussion, we know that such an approximation corresponds
to the condition $v/n\ll 1.$ Similar to the spinless particle case, we can
approximate the Laguerre functions by the Bessel ones to get the following
expression for the radiation intensity 
\begin{equation}
W_{j}=W_{0}\left( \frac{H}{H_{0}}\right) ^{2}(1-\beta _{j}^{2})\sum_{\nu
,j^{\prime }}\int_{0}^{\pi }Q_{jj^{\prime }}^{{\rm cl}}|F_{jj^{\prime }}^{%
{\rm cl}}|^{2}\sin \theta \,d\theta \;.  \label{158}
\end{equation}
The quantities $Q_{jj^{\prime }}^{{\rm cl}}$\ are defined by Eq. (\ref{127})
and $F_{jj^{\prime }}^{{\rm cl}}$ have the form 
\begin{eqnarray}
&&F_{jj^{\prime }}^{{\rm cl}}=\beta \delta _{\zeta ,\zeta ^{\prime
}}F_{jj^{\prime }}^{\left( 0\right) {\rm cl}}+\delta _{\zeta ,-\zeta
^{\prime }}\frac{H}{H_{0}}\frac{1-\beta ^{2}}{2}\bar{\nu}F_{jj^{\prime
}}^{\left( 1\right) {\rm cl}}\,,\;F_{jj^{\prime }}^{\left( 0\right) {\rm cl}%
}=l_{2}I_{jj^{\prime }}^{\prime }(x)+l_{3}\cos \theta \frac{I_{jj^{\prime
}}(x)}{\beta \sin \theta }\,,\;  \nonumber \\
&&F_{jj^{\prime }}^{\left( 1\right) {\rm cl}}=\left( -\zeta \right)
^{j}l_{2}\cos \theta \left[ \frac{I_{jj^{\prime }}(x)}{\beta \sin \theta }%
+\zeta I_{jj^{\prime }}^{\prime }(x)\right] -l_{3}\left[ \frac{%
aI_{jj^{\prime }}(x)}{\beta \sin \theta }+\zeta I_{jj^{\prime }}^{\prime }(x)%
\right] \,,\;  \nonumber \\
&\;&I_{11}(x)=I_{22}(x)=I_{12}(x)=J_{\bar{\nu}}(\bar{\nu}\beta \sin \theta
)\,,\,I_{21}(x)=J_{-\bar{\nu}}(\bar{\nu}\beta \sin \theta )\,,  \nonumber \\
&&x=\bar{\nu}\beta \sin \theta \,,\;a=\cos ^{2}\theta +\sqrt{1-\beta ^{2}}%
\sin ^{2}\theta \,\,.  \label{159}
\end{eqnarray}
In the non-relativistic approximation $\beta ^{2}=2\bar{n}H/H_{0}$, then
results of the previous Section follow from (\ref{158}).

It follows from (\ref{159}) that the solenoid field with $\mu \neq 0$
suppresses the electron self-polarization effect due to transitions $%
j=2\rightarrow j^{\prime }=1$. This suppression can be considered as a
manifestation of AB effect in SR. For $\mu =0$\ such a manifestation
disappears due to the property 
\[
J_{-\nu }(x)=\left( -1\right) ^{\nu }J_{\nu }(x)\,, 
\]
which takes place whenever $\nu $ are integer.

Similarly to the spinless particle case, the degeneracy of the radiation
intensity with respect to the azimuthal quantum number is lifted here
completely. That can be also considered as one of manifestations of AB
effect in SR.

\subsection{Electron transitions from zero energy levels with a change of
state type}

Consider here the radiation intensity caused by electron transitions from $%
n=0$ energy level with a change of the type of state (namely $%
n=0,j=1\rightarrow j^{\prime }=2$ transitions). In this case a superlow
frequency (\ref{72}) is emitted. One can get an exact expression for the
quantity $Q_{12}\,,$\ 
\begin{equation}
Q_{12}=\frac{q^{1-\mu }\exp \left( -q\right) \Phi \left( 1,2-\mu ;q\right) }{%
\Gamma \left( 2-\mu \right) }\,,\;q=\mu \frac{1-\sqrt{p}}{1+\sqrt{p}}%
,\;p=1-\alpha (1-x^{2}),  \label{162}
\end{equation}
where $\Phi (\alpha ,\gamma ;x)$ is the confluent hypergeometric function.
In the case under consideration, we can express $\Phi (\alpha ,\gamma ;x)$
via the incomplete $\Gamma -$function and get the following expression 
\begin{equation}
\Phi \left( 1,2-\mu ;x\right) =\left( 1-\mu \right) x^{\mu
-1}e^{x}\int_{0}^{x}e^{-x}y^{-\mu }dy\,,\;\mu <1\,.  \label{163}
\end{equation}
For the transitions under consideration, the radiation intensity has the
form 
\begin{eqnarray}
&&W=W_{0}\frac{H}{H_{0}}\left( \alpha \mu \right) ^{2}f(\mu )G(\alpha ,\mu
)\,,\;G(\alpha ,\mu )=\int_{0}^{1}\frac{\sqrt{p}+\sqrt{1-\alpha }}{\sqrt{p}%
\left( 1+\sqrt{p}\right) ^{3}}e^{-2q}\Phi \left( 1,2-\mu ;q\right) F\left(
x\right) dx\,,\;  \nonumber \\
\; &&\,F\left( x\right) =\delta _{\zeta ,\zeta ^{\prime }}\frac{1-\zeta }{2}%
\left[ l_{2}^{2}+l_{3}^{2}\psi \left( x\right) \right] +\delta _{\zeta
,-\zeta ^{\prime }}\frac{\alpha }{\left( 1+\sqrt{1-\alpha }\right) ^{2}}%
\frac{1+\zeta }{2}\left[ l_{2}^{2}\psi \left( x\right) +l_{3}^{2}\right] \,,
\nonumber \\
&&\alpha =\frac{2\mu H}{H_{0}+2\mu H}\,,\;\psi \left( x\right) =\frac{%
x^{2}\left( 1+\sqrt{1-\alpha }\right) ^{2}}{\left( \sqrt{p}+\sqrt{1-\alpha }%
\right) ^{2}}\;  \label{164}
\end{eqnarray}
( $f(\mu )$ was defined in (\ref{97})). The function $G(\alpha ,\mu )$\
depends on the magnetic field via the quantity $\alpha ,$ 
\begin{equation}
0<\alpha <1,\;\alpha \approx 2\mu \frac{H}{H_{0}}\;\left( \frac{H}{H_{0}}\ll
1\right) ,\;\lim_{H\rightarrow \infty }\alpha =1\;.  \label{165}
\end{equation}
It is easy to see that 
\begin{equation}
\alpha =1-\left( m/k_{0}\right) ^{2}=\beta ^{2}\,.  \label{166}
\end{equation}
However, for such quantum states ($n=0$), we cannot use a classical
interpretation for $\beta .$

It follows from (\ref{164}) that transitions with and without spin-flip have
almost (with the interchange of $\sigma $ and $\pi $ components) the same
linear polarization of the radiation intensity.

Doing summation over photon polarization states, over final electron spin
states, and averaging over initial spin states, we get total radiation
intensity for a non-polarized electron 
\begin{equation}
\bar{W}=2W_{0}\frac{H}{H_{0}}\alpha \mu ^{2}f(\mu )\int_{0}^{1}\frac{\sqrt{p}%
+\alpha -1}{\sqrt{p}\left( 1+\sqrt{p}\right) ^{3}}e^{-2q}\Phi \left( 1,2-\mu
;q\right) dx\,\,.  \label{167}
\end{equation}
In the weak magnetic field approximation ($\alpha \ll 1$), we obtain from (%
\ref{164}) 
\begin{equation}
W=\frac{1}{3}W_{0}\frac{H}{H_{0}}(\alpha \mu )^{2}f(\mu )\left( \delta
_{\zeta ,\zeta ^{\prime }}\frac{1-\zeta }{2}S_{0}+\delta _{\zeta ,-\zeta
^{\prime }}\frac{\alpha }{4}\frac{1+\zeta }{2}S_{1}\right) \,.  \label{168}
\end{equation}

Finally consider the case of superstrong magnetic fields ($H\gg
H_{0},\;\alpha =1)$). Here $\psi (x)=1$ and the radiation intensity has the
form 
\begin{eqnarray}
W &=&\frac{1}{2}\bar{W}\left( l_{2}^{2}+l_{3}^{2}\right) \left( \delta
_{\zeta ,\zeta ^{\prime }}\frac{1-\zeta }{2}+\delta _{\zeta ,-\zeta ^{\prime
}}\frac{1+\zeta }{2}\right) \,,  \nonumber \\
\bar{W} &=&W_{0}\frac{H}{H_{0}}\mu ^{2}f(\mu )J(\mu ),\;J(\mu
)=\int_{0}^{1}(1+x)e^{-2\mu x}\Phi \left( 1,2-\mu ;\mu x\right) dx\,.
\label{169}
\end{eqnarray}
$J(\mu )$ is a monotonically decreasing function of $\mu .$ In particular,$%
\;J(0)=1,5;\;J(1)=2-\frac{3}{e}\approx 0,896.$ Thus, in the superstrong
magnetic fields, transitions with and without spin-flip have equal
probabilities, the radiation is completely depolarized, and the radiation
intensity is linearly increasing function of the magnetic field.

\section{Summary}

We have obtained exact solutions of Klein-Gordon and Dirac equations in the
magnetic-solenoid field. Employing these solutions, we succeeded to
calculate various characteristics of one-photon radiation in such a field.
Namely, peculiarities of the radiation related to the presence of the AB
solenoid are considered by us as manifestations of AB effect in CR and SR.
Below we list the most important results obtained.

1. It is demonstrated that all the peculiarities of the radiation related to
the presence of AB solenoid depend on the mantissa $\mu $ of the solenoid
flux only. For the fluxes with $\mu =0,$ these peculiarities disappear.

2. The energy spectrum of charge particles in the magnetic-solenoid field
differs essentially from the one in pure magnetic field. In particular, the
degeneracy with respect to the azimuthal quantum number is partially lifted.
Each magnetic field energy level splits in two ones in the magnetic-solenoid
field. In turn, this complicates the radiation spectrum. In particular, the
degeneracy of the radiation intensity with respect to the azimuthal quantum
number is lifted completely.

3. New lines in the radiation spectrum appear, they do not have an analog in
the pure magnetic field case. These lines consist of two series of harmonics
(the latter are not multiple of the basic synchrotron frequency) and of two
superlow frequency harmonics (their frequencies are less than the basic
synchrotron frequency).

4. It is shown that the only one basic synchrotron harmonic and the new
frequencies are irradiated along the magnetic field. We stress important
peculiarities of the radiation along the magnetic field. The basic
synchrotron harmonic has total circular polarization; the radiation
intensity of superlow harmonics has maximum in the magnetic field direction;
all the harmonics from the two above mentioned series have approximately
equal radiation intensities. The latter property of the radiation is not
typical for the conventional CR and SR. We believe that a considerable
relative shift between new harmonics and the basic synchrotron one as well
as the peculiarities of the angular distribution of the radiation intensity
open up possibilities for experimental observation of AB effect in CR and SR.

5. It is discovered that the presence of the solenoid field can suppress the
well-known in SR electron self-polarization effect.

{\bf Acknowledgement} The authors (V.G.B, D.M.G, and A.L) are thankful to
FAPESP for support. (D.M.G) thanks also CNPq for permanent support and
(V.G.B.) thanks Russian Science Ministry Foundation and RFFI for partial
support.

\end{document}